\documentclass[fleqn,usenatbib]{mnras}
\usepackage{newtxtext,newtxmath}
\usepackage[T1]{fontenc}

\usepackage{graphicx}
\usepackage{braket}
\usepackage{bm}
\usepackage{color}
\usepackage{hyperref}
\usepackage{aas_macros}
\usepackage{amsmath}

\title[Cosmological test of local position invariance]{Cosmological test of local position invariance from the asymmetric galaxy clustering}

\author[S. Saga et al.]{
Shohei Saga,$^{1,2}$\thanks{E-mail: saga@iap.fr}
Atsushi Taruya,$^{3,4}$
Yann Rasera$^{5}$
and Michel-Andr{\`e}s Breton$^{6,7,8}$
\\
$^{1}$Sorbonne Universti\'e, CNRS, UMR7095, Institut d'Astrophysique de Paris, 98bis boulevard Arago, F-75014 Paris, France\\
$^{2}$Laboratoire Univers et Th{\'e}ories, Observatoire de Paris, Universit{\'e} PSL, Universit{\'e} de Paris, CNRS, F-92190 Meudon, France\\
$^{3}$Center for Gravitational Physics, Yukawa Institute for Theoretical Physics, Kyoto University, Kyoto 606-8502, Japan\\
$^{4}$Kavli Institute for the Physics and Mathematics of the Universe (WPI), The University of Tokyo Institutes for Advanced Study, The University of Tokyo,\\ 5-1-5 Kashiwanoha, Kashiwa, Chiba 277-8583, Japan\\
$^{5}$Laboratoire Univers et Th{\'e}ories, Universit{\'e} de Paris, Observatoire de Paris, Universit{\'e} PSL, CNRS, F-92190 Meudon, France\\
$^{6}$Aix Marseille Univ, CNRS, CNES, LAM, Marseille, France\\
$^{7}$Institute of Space Sciences (ICE, CSIC), Campus UAB, Carrer de Can Magrans, s/n, 08193 Barcelona, Spain\\
$^{8}$Institut d'Estudis Espacials de Catalunya (IEEC), Carrer Gran Capit\`a 2-4, 08193 Barcelona, Spain
}

\date{Accepted XXX. Received YYY; in original form ZZZ}

\pubyear{2023}

\begin{document}
\label{firstpage}
\pagerange{\pageref{firstpage}--\pageref{lastpage}}
\maketitle

\begin{abstract}
The local position invariance (LPI) is one of the three major pillars of Einstein equivalence principle, ensuring the space-time independence on the outcomes of local experiments. The LPI has been tested by measuring the gravitational redshift effect in various depths of gravitational potentials. We propose a new cosmological test of the LPI by observing the asymmetry in the cross-correlation function between different types of galaxies, which predominantly arises from the gravitational redshift effect induced by the gravitational potential of haloes at which the galaxies reside. We show that the ongoing/upcoming galaxy surveys give a fruitful constraint on the LPI-violating parameter, $\alpha$, in the distant universe (redshift $z\sim0.1$--$1.8$) over the cosmological scales (separation $s\sim5$--$10\, {\rm Mpc}/h$) that have not yet been explored, finding that the expected upper limit on $\alpha$ can reach $0.03$.
\end{abstract}

\begin{keywords}
Cosmology -- large-scale structure of Universe --  dark matter
\end{keywords}



\section{Introduction}
Since its foundation, general relativity has been the essential framework to describe gravity in astronomy and cosmology. An important building block of general relativity is the Einstein equivalence principle. As part of it, the local position invariance (LPI) has been playing a special role even for alternative theories of gravity. It states that the outcome of any local non-gravitational experiment is independent of where and when it is performed.
An important consequence of the LPI is predicting the gravitational redshift effect. It indicates that the gravitational redshift, $z_{\rm grav}$, between two identical clocks located at different gravitational potentials, $\Delta \phi$, can be given by $z_{\rm grav} = \Delta \phi$ (the speed of light is taken as unity).
If the LPI is violated, this relation has to be modified, and it is commonly parametrized in the form \citep[e.g.,][]{will_2018}:
\begin{align}
z_{\rm grav} = (1+\alpha) \Delta \phi , \label{eq: def alpha}
\end{align}
where the non-zero value of $\alpha$ implies the LPI violation.

\citet{1959PhRvL...3..439P,1965PhRv..140..788P} have made the first successful high-precision measurements of the gravitational redshift effect due to the gravitational potential of the Earth (the Pound-Rebka-Snider experiment), constraining the LPI-violating parameter with an accuracy of $\alpha\lesssim O(10^{-2})$.
After these pioneering works, the constraint on $\alpha$ has been obtained and improved by many measurements, for instance, spacecraft measurements~\citep{1980PhRvL..45.2081V,1990PhRvL..64.1322K}, solar-spectra measurements~\citep{1991ApJ...376..757L,2020A&A...643A.146G}, and null experiments, which constrain the difference in $\alpha$ between different kinds of atomic clocks in the laboratory~\citep{2013PhRvL.111f0801L,2013PhRvA..87a0102P}.
Recent null experiment puts the upper bound on the LPI-violating parameter by $\alpha < O(10^{-6})$~\citep{2013PhRvA..87a0102P}.
The limits on $\alpha$ obtained above cover a range of $10^{-15}\lesssim \Delta\phi\lesssim 10^{-6}$.
Interestingly, \citet{2019PhRvL.122j1102A,2021ApJ...914..112M} have recently measured the stellar/quasar spectrum near the galactic centre supermassive black hole and gave a limit on an LPI violation of $\alpha\lesssim 10^{-2}$ with a potential difference $10^{-4}\lesssim \Delta\phi\lesssim 10^{-2}$.

In this paper, we propose a novel cosmological test of the LPI, probing a new region $10^{-6}\lesssim \Delta\phi\lesssim 10^{-4}$, by using the measurements of galaxy redshift surveys (this range is also covered by the surface gravity of stars, but separating the surface gravity from the systematic effects is still difficult e.g., \citealt{2019ApJ...871..119D,2022MNRAS.514.1071M}).
The observed galaxy distributions via spectroscopic measurement are apparently distorted due to the special and general relativistic effects~\citep[e.g.,][]{1987MNRAS.228..653S,2000ApJ...537L..77M,2013MNRAS.434.3008C,2014CQGra..31w4001Y,2018JCAP...03..019T,2009JCAP...11..026M,2014PhRvD..89h3535B}.
Some of the relativistic effects induce asymmetric distortions along the line-of-sight when cross-correlating different types of galaxies, leading to a non-vanishing dipole.

\citet{2018JCAP...05..061B,2020JCAP...08..004B} have pointed out that the large-scale dipole signal can test the weak equivalence principle. In contrast, based on the numerical simulations and analytical model, we have recently shown that the small-scale dipole is dominated by the gravitational redshift effect mainly arising from the gravitational potential of dark matter haloes~\citep{2019MNRAS.483.2671B,2020MNRAS.498..981S}.
Further, we found that such a signal can be detected from upcoming galaxy surveys at a statistically significant level~\citep{2022MNRAS.511.2732S} \citep[see][for a similar forecast based on a different approach]{2020JCAP...07..048B}.
Note that even the current data set of galaxy clustering and clusters of galaxies provide a marginal detection \citep{2017MNRAS.470.2822A,2011Natur.477..567W,2013MNRAS.435.1278K,2015PhRvL.114g1103S,2015MNRAS.448.1999J,2017MNRAS.468.1981C,2021MNRAS.503..669M}.
We thus anticipate that the detected dipole signals from future surveys enable us to measure the gravitational redshift effect, offering the LPI test at cosmological scales.

Motivated by these, we present a quantitative analysis for the forecast constraint on the LPI-violating parameter $\alpha$. In doing so, we use an analytical model that reproduces numerical simulations quite well~\citep{2020MNRAS.498..981S,2022MNRAS.511.2732S}. Taking two major systematics arising from off-centred galaxies into account, we demonstrate that future galaxy surveys will offer an insightful cosmological test of the LPI, uncovering the parameter space that has not been explored so far.

This paper is organized as follows. In Sec.~\ref{sec: model}, we present the model of the dipole moment based on our previous works~\citep{2020MNRAS.498..981S,2022MNRAS.511.2732S}, in which the major relativistic effects, the gravitational redshift and transverse Doppler effects are taken into account. In Sec.~\ref{sec: fisher}, we derive the expected constraint on the LPI-violating parameter $\alpha$ by ongoing and upcoming galaxy redshift surveys. Sec.~\ref{sec: summary} is devoted to summary and discussions. Appendices~\ref{sec: app Eq3}, \ref{sec: app model NL}, and \ref{sec: app Fisher} provide, respectively, the derivation of the analytical model for the dipole, the model of the non-perturbative terms involved in the dipole, and the details of the Fisher analysis for deriving uncertainty in the bias parameters, which is used in obtaining the constraint on $\alpha$.

\section{Relativistic dipole}
\label{sec: model}
Let us recall that the observed galaxy position via spectroscopic surveys receives relativistic corrections through the light propagation in an inhomogeneous universe on top of the cosmic expansion. Consequently, the observed source position, $\bm{s}$, differs generally from the true position, $\bm{x}$. Taking the major effects into account, their relation becomes
\begin{align}
\bm{s} &= \bm{x} + \frac{1}{aH}
\left[
\left( \bm{v}\cdot \hat{\bm{x}}\right)
- \phi_{\rm halo}
+ \gamma v^{2}_{\rm g}
\right]
\hat{\bm{x}} ,
\label{eq: s to x}
\end{align}
with $\hat{\bm x}$ being the unit vector, $\hat{\bm{x}} = \bm{x}/|\bm{x}|$. The quantities $a$, $H$, and $\bm{v}$ are a scale factor, Hubble parameter, and peculiar velocity of galaxies, respectively. The explicit form of other minor contributions are found in e.g., \citet{2010PhRvD..82h3508Y,2011PhRvD..84d3516C,2011PhRvD..84f3505B}.
In Eq.~(\ref{eq: s to x}), three contributions in the square bracket are, from the first to third terms, (i) the longitudinal Doppler effect induced by the galaxy peculiar motion, (ii) gravitational redshift effect arising from the potential of the halo at the galaxy position, $\phi_{\rm halo}$\footnote{We ignore the contribution from the linear density field, which could be important to probe the equivalence principle at large scales. However, such a term produces a negligible gravitational redshift at the scales of interest.}, and finally, (iii) sum of transverse Doppler, light-cone, and surface brightness modulation effects mainly due to the virialized random motion of galaxies, $v^{2}_{\rm g}$~\citep{2013MNRAS.435.1278K,2015MNRAS.448.1999J,2017MNRAS.468.1981C,2021MNRAS.503..669M}. We introduced a parameter $\gamma$ and use $\gamma = -5/2$ as a fiducial value, taken from \citet{2013MNRAS.435.1278K}. We will discuss the impact of the uncertainty in this parameter later.
Since the second and third terms largely depend on the halo properties of targeted galaxies, they are systematically treated as a deterministic constant rather than a stochastic variable, determined solely by the halo masses. In Eq.~(\ref{eq: s to x}), relativistic corrections systematically change the observed position along the specific direction $\hat{\bm{x}}$. This apparently produces an asymmetry in the galaxy clustering, and taking a pair of galaxies with different sizes of relativistic corrections results in a non-vanishing dipole (see Eq.~(\ref{eq: xi1})).

With the mapping relation in Eq.~(\ref{eq: s to x}) and the number conservation, the observed number density fluctuation of the galaxy population X, $\delta^{(\rm S)}_{\rm X}(\bm{s})$ is related to the real-space galaxy density field, $\delta_{\rm X}(\bm{x})$. 
Furthermore, the galaxy distribution is a biased tracer of matter fluctuations. Our assumption here is that it is simply related to the linear matter fluctuations $\delta_{L}(\bm{x})$ through $\delta_{\rm X}(\bm{x}) = b_{\rm X}\delta_{L}(\bm{x})$, with the linear bias parameter $b_{\rm X}$ to be determined observationally.
Treating the density fluctuations and relativistic corrections perturbatively, the quantity $\delta_{\rm X}^{\rm(S)}$ is solely expressed in terms of $\delta_L$ \citep{2022MNRAS.511.2732S}.

We then compute the correlation function between the galaxy populations X at $\bm{s}_{1}$ and Y at $\bm{s}_{2}$, $\xi(\bm{s}_1,\bm{s}_2) \equiv \Braket{\delta^{(\rm S)}_{\rm X}(\bm{s}_{1}) \delta^{(\rm S)}_{\rm Y}(\bm{s}_{2})}$ with $\langle\cdots\rangle$ being the ensemble average. Without loss of generality, we write it as a function of the separation $s = |\bm{s}_{2}-\bm{s}_{1}|$, line-of-sight distance $d = |(\bm{s}_{1}+\bm{s}_{2})/2|$, and directional cosine between the line-of-sight and separation vectors, $\mu = \hat{\bm{s}}\cdot\hat{\bm{d}}$. The dipole moment of the correlation function characterizing the asymmetric galaxy clustering is defined by $\xi_{1}(s,d) = \frac{3}{2}\int^{1}_{-1}{\rm d}\mu\, \mu\, \xi(s,d,\mu)$.
For the scales of $s/({\rm Mpc}/h) \lesssim 30$ in the distant universe, the terms of order $O\left( (s/d)^{2} \right)$ are small, and dropping them, the non-zero contributions to the dipole are given by (see Appendix~\ref{sec: app Eq3})
\begin{align}
\xi_{1}(s,d) &=
2f\Delta b\frac{s}{d}\left( \Xi^{(1)}_{1}(s) - \frac{\Xi^{(0)}_{2}(s)}{5} \right)
+ \left( \Delta\phi + \frac{5}{2} \Delta v^{2}_{\rm g} \right)
\notag \\
& \times
\frac{1}{saH} \left( b_{\rm X}b_{\rm Y} + \frac{3}{5}(b_{\rm X}+b_{\rm Y})f + \frac{3}{7}f^{2}\right) \Xi^{(-1)}_{1}(s)
, \label{eq: xi1}
\end{align}
where the function $f\equiv {\rm d}\ln{D}/{\rm d}\ln{a}$ is the linear growth rate, with $D$ being the linear growth factor.
We define $\Xi^{(n)}_{\ell}(s) \equiv \int k^{2}\, {\rm d}k/(2\pi^{2})\, j_{\ell}(ks) (ks)^{-n} P_{\rm L}(k)$ with $j_{\ell}$ and $P_{\rm L}(k)$ being, respectively, the spherical Bessel function and linear matter power spectrum.
In Eq.~(\ref{eq: xi1}), all terms are proportional to the differential quantities, i.e., $\Delta b\equiv b_{\rm X}-b_{\rm Y}$, $\Delta \phi \equiv \phi_{\rm halo,X}-\phi_{\rm halo,Y}$, and $\Delta v^{2}_{\rm g}\equiv v^{2}_{\rm g, X} - v^{2}_{\rm g, Y}$. Accordingly, the non-vanishing dipole arises only when we cross-correlate different biased objects, ${\rm X} \neq {\rm Y}$. 
To observe the dipole signal and use it as the probe of the LPT test, we preferentially cross-correlate the objects having a similar potential depth, which can be regarded as the conditional average over density fields, $\delta$, for a given potential depth.
Hence, a non-zero contribution from, e.g., $\Braket{\Phi\delta\delta}$, would be present but be small, indeed justified in comparing the analytical predictions with the simulations~\citep{2019MNRAS.483.2671B}.

We note that our model (\ref{eq: xi1}) ignores the magnification bias due to the flux-limited galaxy samples that also contributes to the dipole \citep{2014PhRvD..89h3535B,2017PhRvD..95d3530H}.
As shown in \citet{2022MNRAS.511.2732S}, its impact is small at the scales of interest and hence does not change our results.
We keep the Doppler contribution which is also negligible compared to the second term in Eq.~(\ref{eq: xi1}) at small scales, but has the comparable amplitude to the magnification bias. If we cross-correlate subhaloes or satelite galaxies, the Doppler contribution may become a non-negligible effect as the Finger-of-God effect, although it has not been appropriately modelled beyond the plane-parallel limit yet, which is beyond the scope of this work.

Given the linear matter power spectrum and bias parameters, the remaining pieces to be specified for a quantitative prediction of $\xi_1$ are $\phi_{\rm halo}$ and $v^{2}_{\rm g}$, which are modelled by the universal halo density profile, called Navarro-Frenk-White profile~\citep{1996ApJ...462..563N}.
Assuming its functional form is characterized by halo mass and redshift, the halo potential, $\phi_{\rm halo}$, is obtained by solving the Poisson equation, while the velocity dispersion, $v^{2}_{\rm g}$, is computed from the Jeans equation \citep{2020MNRAS.498..981S,2022MNRAS.511.2732S}.
Here, we also add the halo coherent motion to $v_{\rm g}^2$ according to \citet{2013PhRvD..88d3013Z,2017MNRAS.471.2345Z,2019JCAP...04..050D}. In predicting the dipole, a crucial aspect is that each of the galaxies to cross-correlate does not strictly reside at the halo centre. The presence of the off-centred galaxies induces two competitive effects, i.e., the diminution of the gravitational redshift and non-vanishing transverse Doppler effects, which systematically change the dipole amplitude. We account for them following \citet{2013MNRAS.435.2345H,2020MNRAS.493.1120Y}, and control their potential impact by introducing the off-centring parameter $R_{\rm off}$. As a result, we can write the parameter dependence explicitly as $\phi_{\rm halo, X}(z, M_{\rm X}, R_{\rm off, X})$ and $v^{2}_{\rm g, X}(z, M_{\rm X}, R_{\rm off, X})$~(see Appendix~\ref{sec: app model NL}).

\begin{figure}
\centering
\includegraphics[width=0.99\columnwidth]{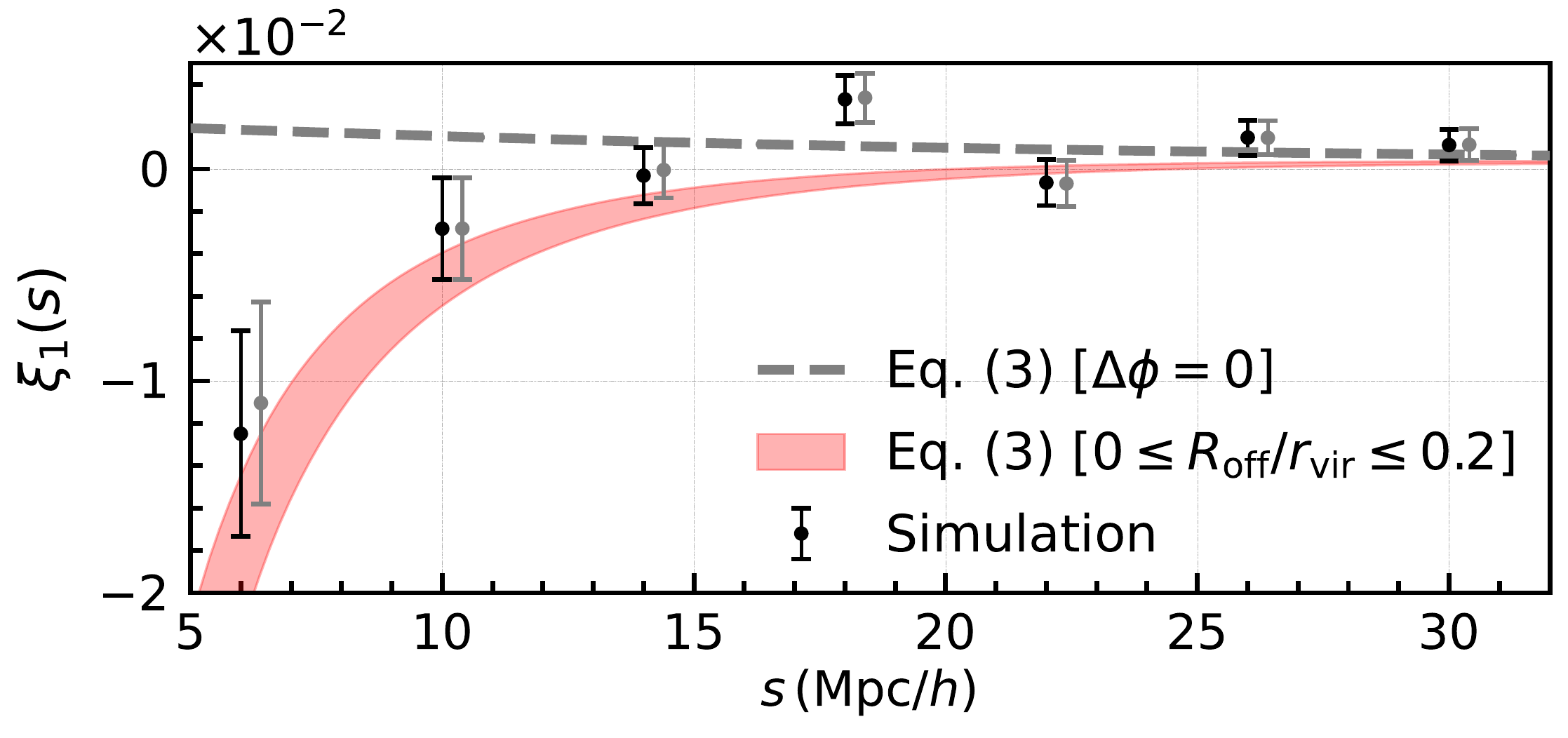}
\caption{Dipole of the correlation function between different biased objects $(b_{\rm X},b_{\rm Y}) = (2.07, 1.08)$ at $z=0.33$ whose halo masses correspond to $(M_{\rm X}, M_{\rm Y}) = (4.4\times 10^{13}\, M_{\odot}/h, 1.6\times 10^{12}\, M_{\odot}/h)$ based on the Sheth-Tormen prescription~\citep{1999MNRAS.308..119S}.
The red shaded curve represents the analytical prediction, varying from $R_{\rm off}/r_{\rm vir}=0$ (lower) to $R_{\rm off}/r_{\rm vir}=0.2$ (upper), leading to the values of the halo potential $(\phi_{\rm halo, X}, \phi_{\rm halo, Y}) = (-1.6\times 10^{-5}, -2.9\times 10^{-6})$ and $(-1.1\times 10^{-5}, -1.7\times 10^{-6})$, respectively, while the grey dashed line ignores the halo potential.
As we measure the dipole of halo-halo cross-correlation in simulations, the values of the velocity dispersion in this analytical prediction are provided based on the large-scale coherent motion alone (see Appendix~\ref{sec: app model NL}), whose values are $(v^{2}_{\rm g,X}, v^{2}_{\rm g,Y}) = (2.2\times 10^{-6}, 2.6\times 10^{-6})$.
The circles with errorbars represent the simulation results taking both Doppler and gravitational redshift effects (black) and all relevant relativistic effects (grey, artificially shifted to the rightward direction for presentation purposes), whose errorbars are estimated using the jack-knife method~\citep{2020MNRAS.498..981S}.
}
\label{fig: dipole}
\end{figure}

Putting all ingredients together, we show the analytical prediction of the dipole at $z=0.33$ in Fig.~\ref{fig: dipole}, together with the measured dipole in simulations incorporating longitudinal Doppler and gravitational redshift effects (black circles), and all the relevant relativistic effects (grey circles)~\citep{2019MNRAS.483.2671B}.
Comparing black with grey circles, the longitudinal Doppler and gravitational redshift effects are shown to be the major contributors to the dipole. Accordingly, it justifies the underlying assumption in our model given in Eq.~(\ref{eq: xi1}).
In this figure, we vary the off-centring parameter by the typical range for the simulations, i.e., the offset of the deepest potential well from the centre of mass position which is the actual halo position defined in simulations: $0 \leq R_{\rm off}/r_{\rm vir}\leq 0.2$ \citep[see e.g.,][]{2020MNRAS.493.1120Y}.
Within the statistical error, the prediction (red curve) describes the simulation results remarkably well down to $5\, {\rm Mpc}/h$, while the prediction ignoring the halo potential (grey dashed) fails to reproduce the negative dipole at small scales.
This suggests that the measurements of the dipole, having particularly a negative amplitude at $s\lesssim 30\,{\rm Mpc}/h$, provide us with information about the gravitational redshift effect from the halo potential, and we can use it to test the LPI violation, as we will see below.

\section{Test of Local Position Invariance}
\label{sec: fisher}
Having confirmed that the analytical predictions properly describe the dipole at the scales of our interest, we next quantitatively consider the prospects for constraining the LPI-violation parameter $\alpha$ in Eq.~(\ref{eq: def alpha}) from upcoming galaxy surveys.

To this end, we perform the Fisher matrix analysis involving several parameters together with $\alpha$ as follows.
\begin{itemize}
    \item {\it Cosmological parameters}: we assume that the cosmological parameters that characterize the linear matter spectrum $P_{\rm L}$ and growth of structure are determined by other cosmological probes e.g., cosmic microwave background (CMB) observations, and fix their fiducial values to the seven-year WMAP results~\citep{2011ApJS..192...18K}.

    \item {\it Bias parameter}: the redshift-space distortions and baryon acoustic oscillations measurements provide the constraint on $b\sigma_{8}$ with $\sigma_{8}$ being the fluctuation amplitude smoothed at $8\, {\rm Mpc}/h$.
    Combining the accurate CMB measurement for power spectrum normalization, we thus have the bias $b$ with a certain error, $\sigma_{b}$ \citep[see e.g.,][]{2003ApJ...598..720S,2011PhRvD..83j3527T}. We obtain the error by performing another Fisher analysis for these observations (see Appendix~\ref{sec: app Fisher} in detail). 
\end{itemize}

On top of these parameters that can be determined independently of the dipole, our theoretical template based on Eq.~(\ref{eq: xi1}) involves parameters associated with the properties of haloes for a given redshift:
\begin{itemize}
    \item {\it Off-centring parameter $R_{\rm off}$}: 
    in principle, we can determine this parameter separately and accurately from the even multipoles \citep[e.g.,][]{2013MNRAS.435.2345H}. Here, we set the typical value, $0.2r_{\rm vir}$, as a fiducial value, and impose a Gaussian prior with the expected errors $\sigma_{R_{\rm off}} = 0.01r_{\rm vir}$, where $r_{\rm vir}$ is the virial radius of haloes \citep[e.g.,][]{2009ApJ...692..217L,2013MNRAS.435.2345H,2020MNRAS.493.1120Y}.
    
    \item {\it Halo mass}: given the bias model described by e.g., the Sheth-Tormen prescription~\citep{1999MNRAS.308..119S}, the halo masses $M_{\rm X/Y}$ are inferred from the bias parameters with errors, $\sigma_{M}= |\partial M/\partial b|\sigma_{b}$. We incorporate this error into our analysis as a Gaussian prior. This treatment enables us to break the degeneracy between the LPI-violating parameter, $\alpha$, and potential difference, $\Delta \phi$ as seen in Eq.~(\ref{eq: def alpha}). The systematic impact of assuming a specific bias model would be reduced, once we can determine the halo masses by complementary probes, e.g., gravitational lensing measurements.
\end{itemize}
To sum up, we have five free parameters in the theoretical template, $\bm{\theta}=\{\alpha$, $R_{\rm off, X/Y}$, and $M_{\rm X/Y}\}$. With the above prescription, the LPI test proposed here is performed consistently under the standard cosmological model.

Let us construct the Fisher matrix. For galaxy samples at the $n$th redshift slice $z_n$, it is given by the $5\times 5$ matrix:
\begin{align}
F_{n,ij} = \sum_{s_{1,2} =s_{\rm min}}^{s_{\rm max}}
\frac{\partial \xi_{1}(s_{1},z_{n})}{\partial \theta_{i}}
\mathcal{C}^{-1}(s_{1},s_{2},z_{n})
\frac{\partial \xi_{1}(s_{2},z_{n})}{\partial \theta_{j}}
, \label{eq: fisher alpha}
\end{align}
with $\mathcal{C}$ being the covariance matrix, which is analytically evaluated by taking only the dominant plane-parallel contributions, ignoring also the non-Gaussian contribution~\citep[][]{2016JCAP...08..021B,2017PhRvD..95d3530H,2022MNRAS.511.2732S}.
We set the minimum separation $s_{\rm min}$ to $5\, {\rm Mpc}/h$, above which the analytical prediction reproduces the simulations, and the systematics of baryonic effects would be negligible. We set $s_{\rm max}$ to $30\, {\rm Mpc}/h$, below which the gravitational redshift effect from the halo potential starts to dominate. Adopting a larger $s_{\rm max}$ hardly changes the results.
Then, with the inverse Fisher matrix at $z_{n}$, $\sigma^{2}_{n,\alpha} \equiv F^{-1}_{n,\alpha\alpha}$, we combine all the redshift bins by $\sigma_{\alpha} = 1/\sqrt{\sum_{n}\sigma^{-2}_{n,\alpha}}$, which gives the expected $1\sigma$ error for a given survey on the LPI-violating parameter, marginalizing over other parameters.

\begin{figure}
\centering
\includegraphics[width=0.99\columnwidth]{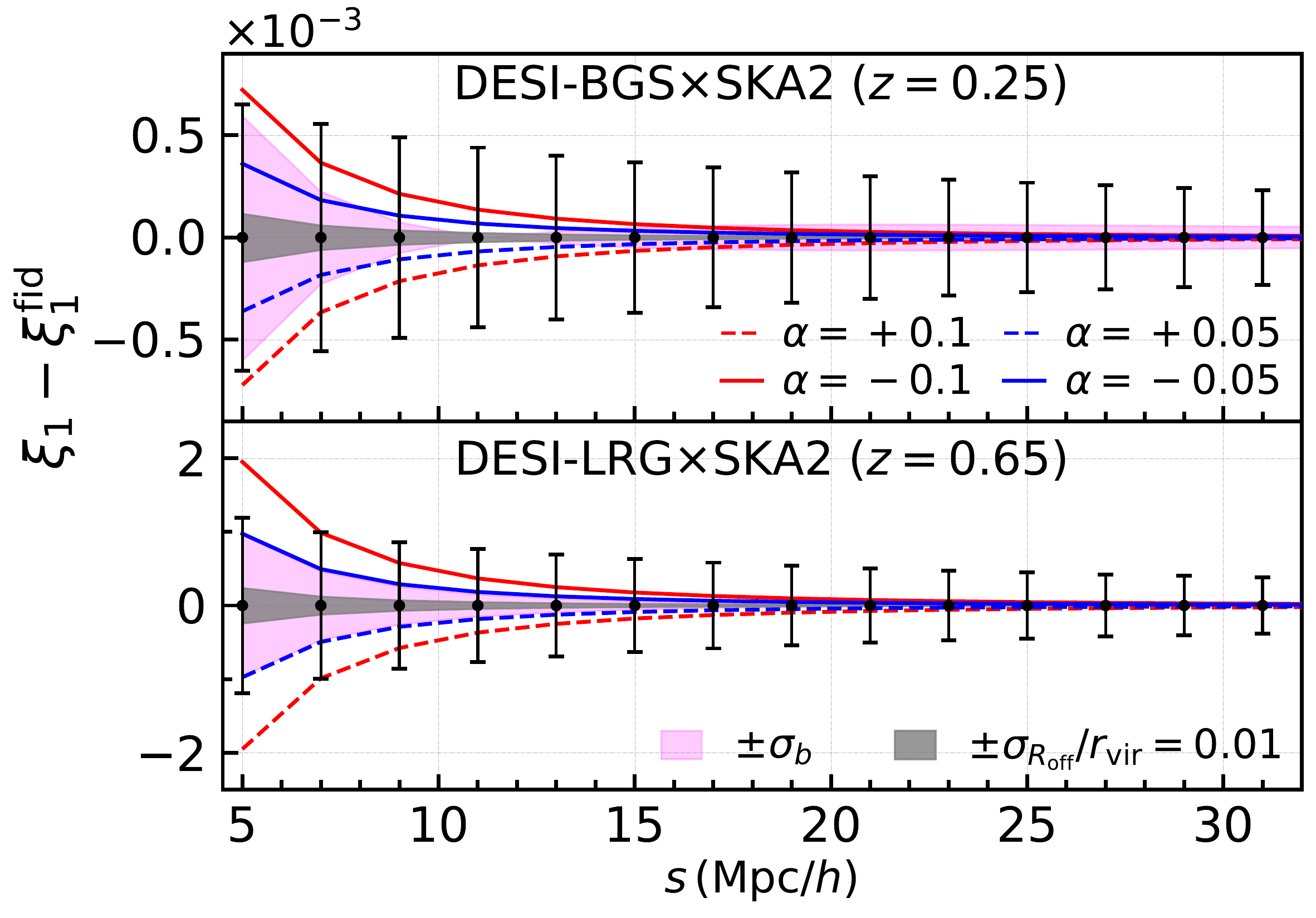}
\caption{Expected $1\sigma$ error on the dipole around the fiducial signal, 
$(R_{\rm off}/r_{\rm vir},\alpha)=(0.2,0)$, for the future surveys as indicated. The red and blue lines represent the difference between the fiducial signal and the one when $\alpha = \pm 0.1$ and $\pm 0.05$, respectively. The grey and magenta regions represent, respectively, the variation of the signal when varying the off-centring parameter and bias parameter within the prior range: $\sigma_{R_{\rm off}}/r_{\rm vir} = 0.01$, $(\sigma_{b,{\rm DESI-BGS}}, \sigma_{b,{\rm SKA2}})=(0.05, 0.02)$ at $z=0.25$, and $(\sigma_{b,{\rm DESI-LRG}}, \sigma_{b,{\rm SKA2}})= (0.04, 0.01)$ at $z=0.65$.
}
\label{fig: xi1-xi1fid}
\end{figure}

Our Fisher matrix analysis considers the cross-correlation function between two distinct galaxy populations obtained from different surveys, assuming that these surveys are maximally overlapped. We examine the combination of the following surveys: Dark Energy Spectroscopic Instrument (DESI) targeting magnitude-limited Bright Galaxies (BGS), Luminous Red Galaxies (LRGs), and Emission Line Galaxies (ELGs)~\citep{2016arXiv161100036D},
Euclid targeting ${\rm H}\alpha$ emitters~\citep{2011arXiv1110.3193L},
Subaru Prime Focus Spectrograph (PFS) targeting ${\rm OII}$ ELGs~\citep{2014PASJ...66R...1T}, and Square Kilometre Array (SKA) targeting ${\rm HI}$ galaxies with two phases dubbed SKA1 and SKA2~\citep{2020PASA...37....7S}
(see Appendix~E of \citealt{2022MNRAS.511.2732S} for the survey parameters).
Note that splitting galaxies obtained from a single survey, a cross correlation between two subsamples would also yield a non-zero dipole. However, its detectability strongly depends on how we split the sample \citep[see][]{2022MNRAS.511.2732S}. In this paper, we rather focus on a solid way that combines two distinct surveys.

Before presenting forecast results, Fig.~\ref{fig: xi1-xi1fid} shows, for illustration, the expected errors around the predicted dipole for DESI-BGS$\times$SKA2 ({\it top}) and DESI-LRG$\times$SKA2 ({\it bottom}), with the fiducial setup (i.e., $R_{\rm off}/r_{\rm vir}=0.2$ and $\alpha=0$) at the specific redshifts, $z=0.25$ and $z=0.65$, respectively. Here, we also show the expected signals when changing the LPI-violating parameter to $\alpha=\pm0.1$ and $\pm 0.05$, varying the off-centring parameter within the prior range $\sigma_{R_{\rm off}}/r_{\rm vir} = 0.01$ (grey), and varying the bias parameter within the prior range $\sigma_{b}$ derived by performing another Fisher analysis (see Appendix~\ref{sec: app Fisher}) as indicated in the caption of the figure (magenta). 
Fig.~\ref{fig: xi1-xi1fid} suggests that the dipole signal from upcoming surveys allows us to measure an LPI violation of the order of $\mathcal{O}(\alpha)=0.01$ even for a single redshift slice if the other parameters are held fixed.

\begin{figure}
\includegraphics[width=0.99\columnwidth]{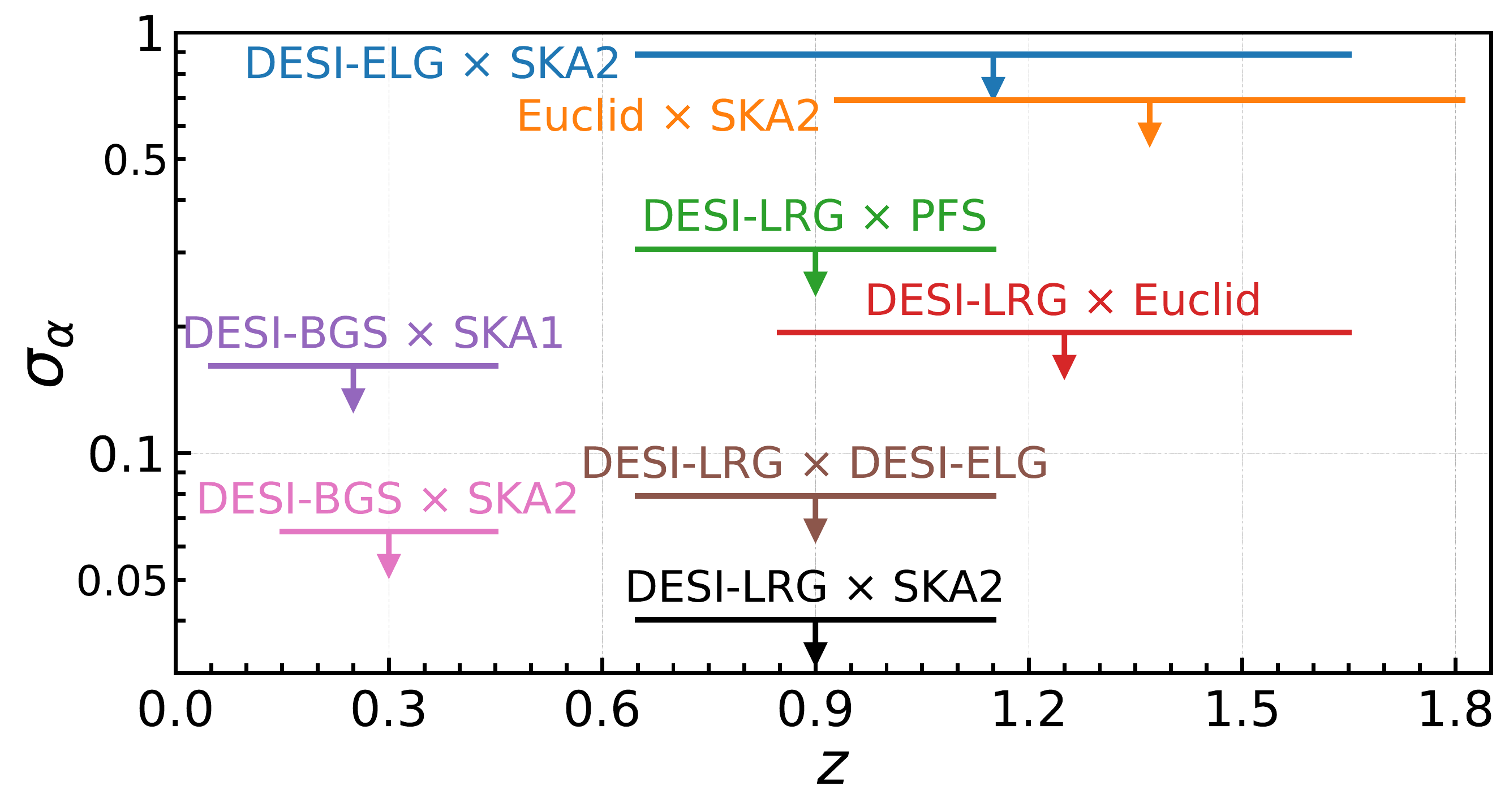}
\caption{Expected $1\sigma$ error on the LPI-violating parameter $\alpha$, obtained by cross-correlating different sources observed in different surveys as indicated. The length of each horizontal line represents the survey redshift ranges to obtain the $1\sigma$ error.}
\label{fig: sigma alpha}
\end{figure}

Computing the Fisher matrix, we obtain the $1\sigma$ error on the LPI-violating parameter with other parameters marginalized over. Fig.~\ref{fig: sigma alpha} shows the results from various combinations of upcoming surveys against the redshift.
Among these, the combination of DESI-LRG and SKA2 gives the tightest constraint with $\sigma_{\alpha} \approx 0.040$ at $0.7 \lesssim z \lesssim 1.1$. This is attributed to the large bias difference, increasing the signal, and a large number of galaxies in SKA2, reducing the shot noise.
Assuming that all the observations shown in Fig.~\ref{fig: sigma alpha} are independent, combining all measurements  further improves the constraint on $\alpha$ down to $\sigma_{\alpha}\approx 0.030$.

\begin{figure*}
\centering
\includegraphics[width=0.99\textwidth]{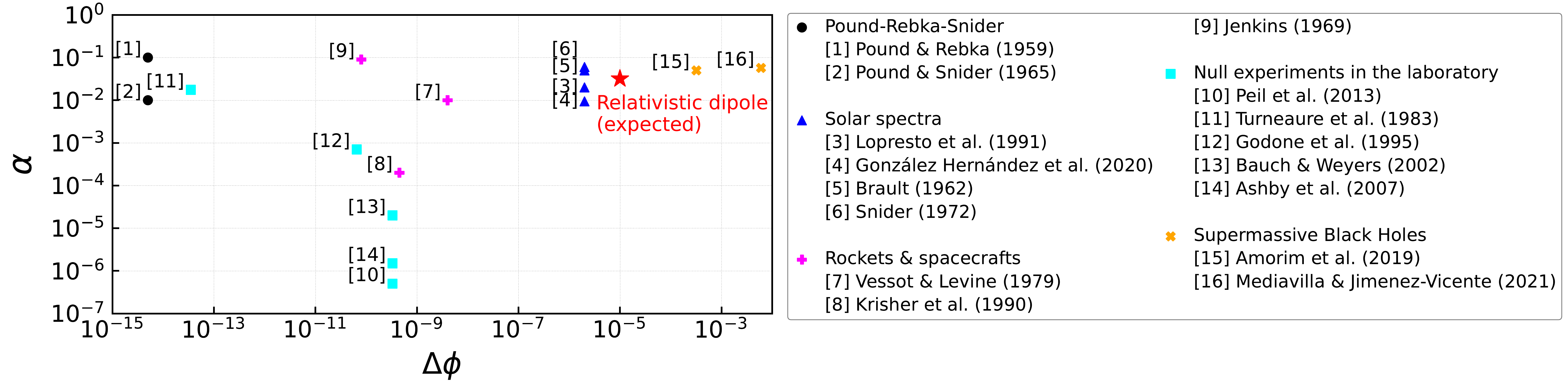}
\caption{
Upper limits on the LPI-violating parameter $\alpha$ as a function of the difference of gravitational potentials by Pound-Rebka-Snider experiments~\citep{1959PhRvL...3..439P,1965PhRv..140..788P}, solar spectra measurements~\citep{1962PhDT........57B,1972PhRvL..28..853S,1991ApJ...376..757L,2020A&A...643A.146G}, rockets and spacecraft experiments~\citep{1969AJ.....74..960J,1979GReGr..10..181V,1990PhRvL..64.1322K}, null experiments~\citep{1983PhRvD..27.1705T,1995PhRvD..51..319G,2002PhRvD..65h1101B,2007PhRvL..98g0802A,2013PhRvA..87a0102P}, and observations of stars/quasars near the galactic centre supermassive black hole~\citep{2019PhRvL.122j1102A,2021ApJ...914..112M} as indicated in the right table.
This figure is based on Fig.~3 of \citet{2019PhRvL.122j1102A}, with our forecast result added (indicated by star symbol).}
\label{fig: LPI limit}
\end{figure*}

The expected upper limit from the proposed LPI test is compared to the previous results for various potential differences $\Delta\phi$, summarized in Fig.~\ref{fig: LPI limit}. In contrast to the previous results, it is worth noting that: (i) the dipole measurement can be a unique probe to explore a new parameter space of the LPI violation, i.e., $\Delta\phi\approx 10^{-5}$, (ii) our method is a new cosmological approach that cannot be categorized as any previous method, and (iii) the method enables us, for the first time, to constrain the LPI violation at cosmological scales.

\section{Summary and discussion}
\label{sec: summary}
In this paper, we have explicitly shown that the cross-correlation function between galaxies with different host haloes and clustering bias yields a non-vanishing dipole.
Such a feature typically appears at $s\lesssim30$\,Mpc$/h$, and is dominated by the gravitational redshift effect from the potential of haloes hosting observed galaxies. Analytical predictions combining perturbation theory with halo model prescription agree well with simulations taking the relativistic effects into account. The Fisher matrix analysis based on the analytical model showed that despite the systematics arising from the off-centred galaxies, the dipole measured from the upcoming galaxy surveys offers a unique LPI test at cosmological scales in the high-redshift universe. While the achievable precision of the LPI-violating parameter, $\alpha\lesssim 0.030$, is comparable to the upper limit from the Pound-Rebka-Snider experiments~\citep{1959PhRvL...3..439P,1965PhRv..140..788P} and is weaker than the recent tests based on the null experiments, the proposed method allows us to probe the potential depth of $\Delta\phi\approx10^{-5}$, which has not been fully explored.

The outcome of our Fisher matrix analysis relies on several simplifications and specific setups. Among these, our theoretical template adopts the halo model prescription assuming the one-to-one correspondence between galaxy and halo. Hence, the predicted amplitude of the dipole signal is tightly linked to the halo mass.
A more careful modelling based on numerical simulations, though not qualitatively affecting the present results, is required for more precision, taking a proper account of the realistic halo-galaxy connection as well as systematic effects from the assembly bias characterizing the secondary halo properties~\citep{2005MNRAS.363L..66G,2007MNRAS.374.1303C}, velocity bias~\citep{2015PhRvD..92l3507B,2019PhRvD.100h3504M}, or different properties between infalling and outward moving galaxies implied by quenching~\citep[e.g.,][]{2005MNRAS.359..949B,2022ApJ...930...43W}. Modelling them is a common challenge when investigating galaxy-galaxy correlations, for which we need a certain model refined by hydrodynamic simulations, and is therefore beyond the scope of this work. However, once the model including the above systematic effects have been developed, our proposed methodology is still available.

Finally, we quantitatively discuss below that meaningful constraints are still possible in more conservative situations. First, we consider the impact of off-centred galaxies.
While our setup of the off-centring parameter and its Gaussian prior is reasonable for LRGs, upcoming surveys will also observe ELGs, whose properties might not necessarily be the same. Nevertheless, even with a conservative choice of $R_{\rm off}/r_{\rm vir} = 0.4$ and a weak prior condition $\sigma_{R_{\rm off}}/r_{\rm vir}=0.1$, the degradation of the constraint on $\alpha$ is found to be moderate, and we can still perform a meaningful test from future surveys, with the LPI-violation parameter constrained to be $\sigma_{\alpha}\approx 0.041$.
Second, we further take $10\%$ uncertainty in $\gamma$, which roughly corresponds to the typical variation range of the spectral index for the galaxy spectral energy distributions~\citep{2013MNRAS.435.1278K} into account in the Fisher analysis, and found the result does not significantly change ($\sigma_{\alpha}\approx 0.032$).
Third, we further examined how the uncertainty in the bias parameter affects the derived constraints on $\alpha$. We perform the same analysis again for the more conservative case by setting the uncertainty in the bias parameter to twice the fiducial value.
Then, we again found that the result does not significantly change~($\sigma_{\alpha} \approx 0.034$).

Since our method provides a consistency test under general relativity, a non-zero detection of the LPI-violating parameter does not simply imply LPI violation, which is generally inherent in the LPI test.
However, we can generalize our methodology to modified gravity, requiring a more systematic and elaborate study, which is beyond the scope of this paper.
A pursuit of measuring the dipole signal is indispensable, and the present method will pave a pathway to the cosmological LPI test in the distant universe.


\section*{Acknowledgements}
This work was initiated during the invitation program of JSPS Grant No. L16519. Numerical simulation was granted access to HPC resources of TGCC through allocations made by GENCI (Grand Equipement National de Calcul Intensif) under the allocations A0030402287, A0050402287, A0070402287 and A0090402287.
Numerical computation was also carried out partly at the Yukawa Institute Computer Facility. This work was supported by Grant-in-Aid for JSPS Fellows No.~17J10553 (SS) and in part by MEXT/JSPS KAKENHI Grant Numbers Nos. JP17H06359, JP20H05861, and 21H01081 (AT). AT also acknowledges the support from JST AIP Acceleration Research Grant No. JP20317829, Japan. SS acknowledges the support from Yukawa Institute for Theoretical Physics (YITP) at Kyoto University, where this work was completed during the visiting program. Also, discussions during the workshop YITP-T-21-06 on ``Galaxy shape statistics and cosmology'' were useful to complete this work.

\section*{Data Availability}
The data underlying this article are available in the article.

\bibliographystyle{mnras}
\bibliography{ref}

\begin{thebibliography}{}
\makeatletter
\relax
\def\mn@urlcharsother{\let\do\@makeother \do\$\do\&\do\#\do\^\do\_\do\%\do\~}
\def\mn@doi{\begingroup\mn@urlcharsother \@ifnextchar [ {\mn@doi@}
  {\mn@doi@[]}}
\def\mn@doi@[#1]#2{\def\@tempa{#1}\ifx\@tempa\@empty \href
  {http://dx.doi.org/#2} {doi:#2}\else \href {http://dx.doi.org/#2} {#1}\fi
  \endgroup}
\def\mn@eprint#1#2{\mn@eprint@#1:#2::\@nil}
\def\mn@eprint@arXiv#1{\href {http://arxiv.org/abs/#1} {{\tt arXiv:#1}}}
\def\mn@eprint@dblp#1{\href {http://dblp.uni-trier.de/rec/bibtex/#1.xml}
  {dblp:#1}}
\def\mn@eprint@#1:#2:#3:#4\@nil{\def\@tempa {#1}\def\@tempb {#2}\def\@tempc
  {#3}\ifx \@tempc \@empty \let \@tempc \@tempb \let \@tempb \@tempa \fi \ifx
  \@tempb \@empty \def\@tempb {arXiv}\fi \@ifundefined
  {mn@eprint@\@tempb}{\@tempb:\@tempc}{\expandafter \expandafter \csname
  mn@eprint@\@tempb\endcsname \expandafter{\@tempc}}}

\bibitem[\protect\citeauthoryear{Aghamousa et~al.}{Aghamousa
  et~al.}{2016}]{2016arXiv161100036D}
Aghamousa A.,  et~al., 2016, arXiv e-prints, \href
  {https://ui.adsabs.harvard.edu/abs/2016arXiv161100036D} {p. arXiv:1611.00036}

\bibitem[\protect\citeauthoryear{{Alam}, {Zhu}, {Croft}, {Ho}, {Giusarma}  \&
  {Schneider}}{{Alam} et~al.}{2017}]{2017MNRAS.470.2822A}
{Alam} S.,  {Zhu} H.,  {Croft} R. A.~C.,  {Ho} S.,  {Giusarma} E.,
  {Schneider} D.~P.,  2017, \mn@doi [\mnras] {10.1093/mnras/stx1421}, \href
  {https://ui.adsabs.harvard.edu/abs/2017MNRAS.470.2822A} {470, 2822}

\bibitem[\protect\citeauthoryear{{Alcock} \& {Paczynski}}{{Alcock} \&
  {Paczynski}}{1979}]{1979Natur.281..358A}
{Alcock} C.,  {Paczynski} B.,  1979, \mn@doi [\nat] {10.1038/281358a0}, \href
  {https://ui.adsabs.harvard.edu/abs/1979Natur.281..358A} {281, 358}

\bibitem[\protect\citeauthoryear{Amorim et~al.}{Amorim
  et~al.}{2019}]{2019PhRvL.122j1102A}
Amorim A.,  et~al., 2019, \mn@doi [Phys. Rev. Lett.]
  {10.1103/PhysRevLett.122.101102}, 122, 101102

\bibitem[\protect\citeauthoryear{{Ashby}, {Heavner}, {Jefferts}, {Parker},
  {Radnaev}  \& {Dudin}}{{Ashby} et~al.}{2007}]{2007PhRvL..98g0802A}
{Ashby} N.,  {Heavner} T.~P.,  {Jefferts} S.~R.,  {Parker} T.~E.,  {Radnaev}
  A.~G.,   {Dudin} Y.~O.,  2007, \mn@doi [\prl]
  {10.1103/PhysRevLett.98.070802}, \href
  {https://ui.adsabs.harvard.edu/abs/2007PhRvL..98g0802A} {98, 070802}

\bibitem[\protect\citeauthoryear{Bacon et~al.}{Bacon
  et~al.}{2020}]{2020PASA...37....7S}
Bacon D.~J.,  et~al., 2020, \mn@doi [\pasa] {10.1017/pasa.2019.51}, \href
  {https://ui.adsabs.harvard.edu/abs/2020PASA...37....7S} {37, e007}

\bibitem[\protect\citeauthoryear{{Baldauf}, {Desjacques}  \&
  {Seljak}}{{Baldauf} et~al.}{2015}]{2015PhRvD..92l3507B}
{Baldauf} T.,  {Desjacques} V.,   {Seljak} U.,  2015, \mn@doi [\prd]
  {10.1103/PhysRevD.92.123507}, \href
  {https://ui.adsabs.harvard.edu/abs/2015PhRvD..92l3507B} {92, 123507}

\bibitem[\protect\citeauthoryear{{Bardeen}, {Bond}, {Kaiser}  \&
  {Szalay}}{{Bardeen} et~al.}{1986}]{1986ApJ...304...15B}
{Bardeen} J.~M.,  {Bond} J.~R.,  {Kaiser} N.,   {Szalay} A.~S.,  1986, \mn@doi
  [\apj] {10.1086/164143}, \href
  {https://ui.adsabs.harvard.edu/abs/1986ApJ...304...15B} {304, 15}

\bibitem[\protect\citeauthoryear{{Bauch} \& {Weyers}}{{Bauch} \&
  {Weyers}}{2002}]{2002PhRvD..65h1101B}
{Bauch} A.,  {Weyers} S.,  2002, \mn@doi [\prd] {10.1103/PhysRevD.65.081101},
  \href {https://ui.adsabs.harvard.edu/abs/2002PhRvD..65h1101B} {65, 081101}

\bibitem[\protect\citeauthoryear{{Bekki}, {Couch}, {Shioya}  \&
  {Vazdekis}}{{Bekki} et~al.}{2005}]{2005MNRAS.359..949B}
{Bekki} K.,  {Couch} W.~J.,  {Shioya} Y.,   {Vazdekis} A.,  2005, \mn@doi
  [\mnras] {10.1111/j.1365-2966.2005.08932.x}, \href
  {https://ui.adsabs.harvard.edu/abs/2005MNRAS.359..949B} {359, 949}

\bibitem[\protect\citeauthoryear{{Bertacca}, {Maartens}  \&
  {Clarkson}}{{Bertacca} et~al.}{2014}]{2014JCAP...11..013B}
{Bertacca} D.,  {Maartens} R.,   {Clarkson} C.,  2014, \mn@doi [\jcap]
  {10.1088/1475-7516/2014/11/013}, \href
  {https://ui.adsabs.harvard.edu/abs/2014JCAP...11..013B} {2014, 013}

\bibitem[\protect\citeauthoryear{{Beutler} \& {Di Dio}}{{Beutler} \& {Di
  Dio}}{2020}]{2020JCAP...07..048B}
{Beutler} F.,  {Di Dio} E.,  2020, \mn@doi [\jcap]
  {10.1088/1475-7516/2020/07/048}, \href
  {https://ui.adsabs.harvard.edu/abs/2020JCAP...07..048B} {2020, 048}

\bibitem[\protect\citeauthoryear{{Bonvin} \& {Durrer}}{{Bonvin} \&
  {Durrer}}{2011}]{2011PhRvD..84f3505B}
{Bonvin} C.,  {Durrer} R.,  2011, \mn@doi [\prd] {10.1103/PhysRevD.84.063505},
  \href {https://ui.adsabs.harvard.edu/abs/2011PhRvD..84f3505B} {84, 063505}

\bibitem[\protect\citeauthoryear{{Bonvin} \& {Fleury}}{{Bonvin} \&
  {Fleury}}{2018}]{2018JCAP...05..061B}
{Bonvin} C.,  {Fleury} P.,  2018, \mn@doi [\jcap]
  {10.1088/1475-7516/2018/05/061}, \href
  {https://ui.adsabs.harvard.edu/abs/2018JCAP...05..061B} {2018, 061}

\bibitem[\protect\citeauthoryear{{Bonvin}, {Hui}  \& {Gazta{\~n}aga}}{{Bonvin}
  et~al.}{2014}]{2014PhRvD..89h3535B}
{Bonvin} C.,  {Hui} L.,   {Gazta{\~n}aga} E.,  2014, \mn@doi [\prd]
  {10.1103/PhysRevD.89.083535}, \href
  {https://ui.adsabs.harvard.edu/abs/2014PhRvD..89h3535B} {89, 083535}

\bibitem[\protect\citeauthoryear{{Bonvin}, {Hui}  \& {Gaztanaga}}{{Bonvin}
  et~al.}{2016}]{2016JCAP...08..021B}
{Bonvin} C.,  {Hui} L.,   {Gaztanaga} E.,  2016, \mn@doi [\jcap]
  {10.1088/1475-7516/2016/08/021}, \href
  {https://ui.adsabs.harvard.edu/abs/2016JCAP...08..021B} {2016, 021}

\bibitem[\protect\citeauthoryear{{Bonvin}, {Oliveira Franco}  \&
  {Fleury}}{{Bonvin} et~al.}{2020}]{2020JCAP...08..004B}
{Bonvin} C.,  {Oliveira Franco} F.,   {Fleury} P.,  2020, \mn@doi [\jcap]
  {10.1088/1475-7516/2020/08/004}, \href
  {https://ui.adsabs.harvard.edu/abs/2020JCAP...08..004B} {2020, 004}

\bibitem[\protect\citeauthoryear{{Brault}}{{Brault}}{1962}]{1962PhDT........57B}
{Brault} J.~W.,  1962, PhD thesis, PRINCETON UNIVERSITY.

\bibitem[\protect\citeauthoryear{{Breton}, {Rasera}, {Taruya}, {Lacombe}  \&
  {Saga}}{{Breton} et~al.}{2019}]{2019MNRAS.483.2671B}
{Breton} M.-A.,  {Rasera} Y.,  {Taruya} A.,  {Lacombe} O.,   {Saga} S.,  2019,
  \mn@doi [\mnras] {10.1093/mnras/sty3206}, \href
  {https://ui.adsabs.harvard.edu/\#abs/2019MNRAS.483.2671B} {483, 2671}

\bibitem[\protect\citeauthoryear{{Bryan} \& {Norman}}{{Bryan} \&
  {Norman}}{1998}]{1998ApJ...495...80B}
{Bryan} G.~L.,  {Norman} M.~L.,  1998, \mn@doi [\apj] {10.1086/305262}, \href
  {https://ui.adsabs.harvard.edu/abs/1998ApJ...495...80B} {495, 80}

\bibitem[\protect\citeauthoryear{{Bullock}, {Kolatt}, {Sigad}, {Somerville},
  {Kravtsov}, {Klypin}, {Primack}  \& {Dekel}}{{Bullock}
  et~al.}{2001}]{2001MNRAS.321..559B}
{Bullock} J.~S.,  {Kolatt} T.~S.,  {Sigad} Y.,  {Somerville} R.~S.,  {Kravtsov}
  A.~V.,  {Klypin} A.~A.,  {Primack} J.~R.,   {Dekel} A.,  2001, \mn@doi
  [\mnras] {10.1046/j.1365-8711.2001.04068.x}, \href
  {https://ui.adsabs.harvard.edu/abs/2001MNRAS.321..559B} {321, 559}

\bibitem[\protect\citeauthoryear{{Cai}, {Kaiser}, {Cole}  \& {Frenk}}{{Cai}
  et~al.}{2017}]{2017MNRAS.468.1981C}
{Cai} Y.-C.,  {Kaiser} N.,  {Cole} S.,   {Frenk} C.,  2017, \mn@doi [\mnras]
  {10.1093/mnras/stx469}, \href
  {https://ui.adsabs.harvard.edu/abs/2017MNRAS.468.1981C} {468, 1981}

\bibitem[\protect\citeauthoryear{{Challinor} \& {Lewis}}{{Challinor} \&
  {Lewis}}{2011}]{2011PhRvD..84d3516C}
{Challinor} A.,  {Lewis} A.,  2011, \mn@doi [\prd]
  {10.1103/PhysRevD.84.043516}, \href
  {https://ui.adsabs.harvard.edu/abs/2011PhRvD..84d3516C} {84, 043516}

\bibitem[\protect\citeauthoryear{{Cooray} \& {Sheth}}{{Cooray} \&
  {Sheth}}{2002}]{2002PhR...372....1C}
{Cooray} A.,  {Sheth} R.,  2002, \mn@doi [\physrep]
  {10.1016/S0370-1573(02)00276-4}, \href
  {https://ui.adsabs.harvard.edu/abs/2002PhR...372....1C} {372, 1}

\bibitem[\protect\citeauthoryear{{Croft}}{{Croft}}{2013}]{2013MNRAS.434.3008C}
{Croft} R. A.~C.,  2013, \mn@doi [\mnras] {10.1093/mnras/stt1223}, \href
  {https://ui.adsabs.harvard.edu/abs/2013MNRAS.434.3008C} {434, 3008}

\bibitem[\protect\citeauthoryear{{Croton}, {Gao}  \& {White}}{{Croton}
  et~al.}{2007}]{2007MNRAS.374.1303C}
{Croton} D.~J.,  {Gao} L.,   {White} S. D.~M.,  2007, \mn@doi [\mnras]
  {10.1111/j.1365-2966.2006.11230.x}, \href
  {https://ui.adsabs.harvard.edu/abs/2007MNRAS.374.1303C} {374, 1303}

\bibitem[\protect\citeauthoryear{{Dai}, {Li}  \& {Stojkovic}}{{Dai}
  et~al.}{2019}]{2019ApJ...871..119D}
{Dai} D.-C.,  {Li} Z.,   {Stojkovic} D.,  2019, \mn@doi [\apj]
  {10.3847/1538-4357/aaf6aa}, \href
  {https://ui.adsabs.harvard.edu/abs/2019ApJ...871..119D} {871, 119}

\bibitem[\protect\citeauthoryear{{Di Dio} \& {Seljak}}{{Di Dio} \&
  {Seljak}}{2019}]{2019JCAP...04..050D}
{Di Dio} E.,  {Seljak} U.,  2019, \mn@doi [\jcap]
  {10.1088/1475-7516/2019/04/050}, \href
  {https://ui.adsabs.harvard.edu/abs/2019JCAP...04..050D} {2019, 050}

\bibitem[\protect\citeauthoryear{{Di Dio}, {Durrer}, {Marozzi}  \&
  {Montanari}}{{Di Dio} et~al.}{2014}]{2014JCAP...12..017D}
{Di Dio} E.,  {Durrer} R.,  {Marozzi} G.,   {Montanari} F.,  2014, \mn@doi
  [\jcap] {10.1088/1475-7516/2014/12/017}, \href
  {https://ui.adsabs.harvard.edu/abs/2014JCAP...12..017D} {2014, 017}

\bibitem[\protect\citeauthoryear{{Gao}, {Springel}  \& {White}}{{Gao}
  et~al.}{2005}]{2005MNRAS.363L..66G}
{Gao} L.,  {Springel} V.,   {White} S. D.~M.,  2005, \mn@doi [\mnras]
  {10.1111/j.1745-3933.2005.00084.x}, \href
  {https://ui.adsabs.harvard.edu/abs/2005MNRAS.363L..66G} {363, L66}

\bibitem[\protect\citeauthoryear{{Godone}, {Novero}  \& {Tavella}}{{Godone}
  et~al.}{1995}]{1995PhRvD..51..319G}
{Godone} A.,  {Novero} C.,   {Tavella} P.,  1995, \mn@doi [\prd]
  {10.1103/PhysRevD.51.319}, \href
  {https://ui.adsabs.harvard.edu/abs/1995PhRvD..51..319G} {51, 319}

\bibitem[\protect\citeauthoryear{{Gonz{\'a}lez Hern{\'a}ndez}
  et~al.,}{{Gonz{\'a}lez Hern{\'a}ndez} et~al.}{2020}]{2020A&A...643A.146G}
{Gonz{\'a}lez Hern{\'a}ndez} J.~I.,  et~al., 2020, \mn@doi [\aap]
  {10.1051/0004-6361/202038937}, \href
  {https://ui.adsabs.harvard.edu/abs/2020A&A...643A.146G} {643, A146}

\bibitem[\protect\citeauthoryear{{Hall} \& {Bonvin}}{{Hall} \&
  {Bonvin}}{2017}]{2017PhRvD..95d3530H}
{Hall} A.,  {Bonvin} C.,  2017, \mn@doi [\prd] {10.1103/PhysRevD.95.043530},
  \href {https://ui.adsabs.harvard.edu/abs/2017PhRvD..95d3530H} {95, 043530}

\bibitem[\protect\citeauthoryear{{Hamilton}}{{Hamilton}}{1992}]{1992ApJ...385L...5H}
{Hamilton} A.~J.~S.,  1992, \mn@doi [\apjl] {10.1086/186264}, \href
  {https://ui.adsabs.harvard.edu/abs/1992ApJ...385L...5H} {385, L5}

\bibitem[\protect\citeauthoryear{{Hikage}, {Mandelbaum}, {Takada}  \&
  {Spergel}}{{Hikage} et~al.}{2013}]{2013MNRAS.435.2345H}
{Hikage} C.,  {Mandelbaum} R.,  {Takada} M.,   {Spergel} D.~N.,  2013, \mn@doi
  [\mnras] {10.1093/mnras/stt1446}, \href
  {https://ui.adsabs.harvard.edu/abs/2013MNRAS.435.2345H} {435, 2345}

\bibitem[\protect\citeauthoryear{{Jenkins}}{{Jenkins}}{1969}]{1969AJ.....74..960J}
{Jenkins} R.~E.,  1969, \mn@doi [\aj] {10.1086/110889}, \href
  {https://ui.adsabs.harvard.edu/abs/1969AJ.....74..960J} {74, 960}

\bibitem[\protect\citeauthoryear{{Jimeno}, {Broadhurst}, {Coupon}, {Umetsu}  \&
  {Lazkoz}}{{Jimeno} et~al.}{2015}]{2015MNRAS.448.1999J}
{Jimeno} P.,  {Broadhurst} T.,  {Coupon} J.,  {Umetsu} K.,   {Lazkoz} R.,
  2015, \mn@doi [\mnras] {10.1093/mnras/stv117}, \href
  {https://ui.adsabs.harvard.edu/abs/2015MNRAS.448.1999J} {448, 1999}

\bibitem[\protect\citeauthoryear{{Kaiser}}{{Kaiser}}{1987}]{1987MNRAS.227....1K}
{Kaiser} N.,  1987, \mn@doi [\mnras] {10.1093/mnras/227.1.1}, \href
  {https://ui.adsabs.harvard.edu/abs/1987MNRAS.227....1K} {227, 1}

\bibitem[\protect\citeauthoryear{{Kaiser}}{{Kaiser}}{2013}]{2013MNRAS.435.1278K}
{Kaiser} N.,  2013, \mn@doi [\mnras] {10.1093/mnras/stt1370}, \href
  {https://ui.adsabs.harvard.edu/abs/2013MNRAS.435.1278K} {435, 1278}

\bibitem[\protect\citeauthoryear{{Komatsu} et~al.,}{{Komatsu}
  et~al.}{2011}]{2011ApJS..192...18K}
{Komatsu} E.,  et~al., 2011, \mn@doi [\apjs] {10.1088/0067-0049/192/2/18},
  \href {https://ui.adsabs.harvard.edu/abs/2011ApJS..192...18K} {192, 18}

\bibitem[\protect\citeauthoryear{{Krisher}, {Anderson}  \&
  {Campbell}}{{Krisher} et~al.}{1990}]{1990PhRvL..64.1322K}
{Krisher} T.~P.,  {Anderson} J.~D.,   {Campbell} J.~K.,  1990, \mn@doi [\prl]
  {10.1103/PhysRevLett.64.1322}, \href
  {https://ui.adsabs.harvard.edu/abs/1990PhRvL..64.1322K} {64, 1322}

\bibitem[\protect\citeauthoryear{{Laureijs} et~al.}{{Laureijs}
  et~al.}{2011}]{2011arXiv1110.3193L}
{Laureijs} R.,  et~al., 2011, arXiv e-prints, \href
  {https://ui.adsabs.harvard.edu/abs/2011arXiv1110.3193L} {p. arXiv:1110.3193}

\bibitem[\protect\citeauthoryear{{Leefer}, {Weber}, {Cing{\"o}z}, {Torgerson}
  \& {Budker}}{{Leefer} et~al.}{2013}]{2013PhRvL.111f0801L}
{Leefer} N.,  {Weber} C.~T.~M.,  {Cing{\"o}z} A.,  {Torgerson} J.~R.,
  {Budker} D.,  2013, \mn@doi [\prl] {10.1103/PhysRevLett.111.060801}, \href
  {https://ui.adsabs.harvard.edu/abs/2013PhRvL.111f0801L} {111, 060801}

\bibitem[\protect\citeauthoryear{{{\L}okas} \& {Mamon}}{{{\L}okas} \&
  {Mamon}}{2001}]{2001MNRAS.321..155L}
{{\L}okas} E.~L.,  {Mamon} G.~A.,  2001, \mn@doi [\mnras]
  {10.1046/j.1365-8711.2001.04007.x}, \href
  {https://ui.adsabs.harvard.edu/abs/2001MNRAS.321..155L} {321, 155}

\bibitem[\protect\citeauthoryear{{Lopresto}, {Schrader}  \&
  {Pierce}}{{Lopresto} et~al.}{1991}]{1991ApJ...376..757L}
{Lopresto} J.~C.,  {Schrader} C.,   {Pierce} A.~K.,  1991, \mn@doi [\apj]
  {10.1086/170323}, \href
  {https://ui.adsabs.harvard.edu/abs/1991ApJ...376..757L} {376, 757}

\bibitem[\protect\citeauthoryear{{Luki{\'c}}, {Reed}, {Habib}  \&
  {Heitmann}}{{Luki{\'c}} et~al.}{2009}]{2009ApJ...692..217L}
{Luki{\'c}} Z.,  {Reed} D.,  {Habib} S.,   {Heitmann} K.,  2009, \mn@doi [\apj]
  {10.1088/0004-637X/692/1/217}, \href
  {https://ui.adsabs.harvard.edu/abs/2009ApJ...692..217L} {692, 217}

\bibitem[\protect\citeauthoryear{{Matsubara}}{{Matsubara}}{2000}]{2000ApJ...537L..77M}
{Matsubara} T.,  2000, \mn@doi [\apjl] {10.1086/312762}, \href
  {https://ui.adsabs.harvard.edu/abs/2000ApJ...537L..77M} {537, L77}

\bibitem[\protect\citeauthoryear{{Matsubara}}{{Matsubara}}{2019}]{2019PhRvD.100h3504M}
{Matsubara} T.,  2019, \mn@doi [\prd] {10.1103/PhysRevD.100.083504}, \href
  {https://ui.adsabs.harvard.edu/abs/2019PhRvD.100h3504M} {100, 083504}

\bibitem[\protect\citeauthoryear{{McDonald}}{{McDonald}}{2009}]{2009JCAP...11..026M}
{McDonald} P.,  2009, \mn@doi [Journal of Cosmology and Astro-Particle Physics]
  {10.1088/1475-7516/2009/11/026}, \href
  {https://ui.adsabs.harvard.edu/\#abs/2009JCAP...11..026M} {2009, 026}

\bibitem[\protect\citeauthoryear{{Mediavilla} \&
  {Jim{\'e}nez-Vicente}}{{Mediavilla} \&
  {Jim{\'e}nez-Vicente}}{2021}]{2021ApJ...914..112M}
{Mediavilla} E.,  {Jim{\'e}nez-Vicente} J.,  2021, \mn@doi [\apj]
  {10.3847/1538-4357/abfb70}, \href
  {https://ui.adsabs.harvard.edu/abs/2021ApJ...914..112M} {914, 112}

\bibitem[\protect\citeauthoryear{{Moschella}, {Slone}, {Dror}, {Cantiello}  \&
  {Perets}}{{Moschella} et~al.}{2022}]{2022MNRAS.514.1071M}
{Moschella} M.,  {Slone} O.,  {Dror} J.~A.,  {Cantiello} M.,   {Perets} H.~B.,
  2022, \mn@doi [\mnras] {10.1093/mnras/stac1427}, \href
  {https://ui.adsabs.harvard.edu/abs/2022MNRAS.514.1071M} {514, 1071}

\bibitem[\protect\citeauthoryear{{Mpetha} et~al.,}{{Mpetha}
  et~al.}{2021}]{2021MNRAS.503..669M}
{Mpetha} C.~T.,  et~al., 2021, \mn@doi [\mnras] {10.1093/mnras/stab544}, \href
  {https://ui.adsabs.harvard.edu/abs/2021MNRAS.503..669M} {503, 669}

\bibitem[\protect\citeauthoryear{{Navarro}, {Frenk}  \& {White}}{{Navarro}
  et~al.}{1996}]{1996ApJ...462..563N}
{Navarro} J.~F.,  {Frenk} C.~S.,   {White} S. D.~M.,  1996, \mn@doi [\apj]
  {10.1086/177173}, \href
  {https://ui.adsabs.harvard.edu/abs/1996ApJ...462..563N} {462, 563}

\bibitem[\protect\citeauthoryear{{Novikov}}{{Novikov}}{1969}]{1969JETP...30..512N}
{Novikov} E.~A.,  1969, Soviet Journal of Experimental and Theoretical Physics,
  \href {https://ui.adsabs.harvard.edu/abs/1969JETP...30..512N} {30, 512}

\bibitem[\protect\citeauthoryear{{Peil}, {Crane}, {Hanssen}, {Swanson}  \&
  {Ekstrom}}{{Peil} et~al.}{2013}]{2013PhRvA..87a0102P}
{Peil} S.,  {Crane} S.,  {Hanssen} J.~L.,  {Swanson} T.~B.,   {Ekstrom} C.~R.,
  2013, \mn@doi [\pra] {10.1103/PhysRevA.87.010102}, \href
  {https://ui.adsabs.harvard.edu/abs/2013PhRvA..87a0102P} {87, 010102}

\bibitem[\protect\citeauthoryear{{Pound} \& {Rebka}}{{Pound} \&
  {Rebka}}{1959}]{1959PhRvL...3..439P}
{Pound} R.~V.,  {Rebka} G.~A.,  1959, \mn@doi [\prl]
  {10.1103/PhysRevLett.3.439}, \href
  {https://ui.adsabs.harvard.edu/abs/1959PhRvL...3..439P} {3, 439}

\bibitem[\protect\citeauthoryear{{Pound} \& {Snider}}{{Pound} \&
  {Snider}}{1965}]{1965PhRv..140..788P}
{Pound} R.~V.,  {Snider} J.~L.,  1965, \mn@doi [Physical Review]
  {10.1103/PhysRev.140.B788}, \href
  {https://ui.adsabs.harvard.edu/abs/1965PhRv..140..788P} {140, 788}

\bibitem[\protect\citeauthoryear{{Sadeh}, {Feng}  \& {Lahav}}{{Sadeh}
  et~al.}{2015}]{2015PhRvL.114g1103S}
{Sadeh} I.,  {Feng} L.~L.,   {Lahav} O.,  2015, \mn@doi [\prl]
  {10.1103/PhysRevLett.114.071103}, \href
  {https://ui.adsabs.harvard.edu/abs/2015PhRvL.114g1103S} {114, 071103}

\bibitem[\protect\citeauthoryear{{Saga}, {Taruya}, {Breton}  \&
  {Rasera}}{{Saga} et~al.}{2020}]{2020MNRAS.498..981S}
{Saga} S.,  {Taruya} A.,  {Breton} M.-A.,   {Rasera} Y.,  2020, \mn@doi
  [\mnras] {10.1093/mnras/staa2232}, \href
  {https://ui.adsabs.harvard.edu/abs/2020MNRAS.498..981S} {498, 981}

\bibitem[\protect\citeauthoryear{{Saga}, {Taruya}, {Rasera}  \&
  {Breton}}{{Saga} et~al.}{2022}]{2022MNRAS.511.2732S}
{Saga} S.,  {Taruya} A.,  {Rasera} Y.,   {Breton} M.-A.,  2022, \mn@doi
  [\mnras] {10.1093/mnras/stac186}, \href
  {https://ui.adsabs.harvard.edu/abs/2022MNRAS.511.2732S} {511, 2732}

\bibitem[\protect\citeauthoryear{{Sasaki}}{{Sasaki}}{1987}]{1987MNRAS.228..653S}
{Sasaki} M.,  1987, \mn@doi [\mnras] {10.1093/mnras/228.3.653}, \href
  {https://ui.adsabs.harvard.edu/abs/1987MNRAS.228..653S} {228, 653}

\bibitem[\protect\citeauthoryear{{Seo} \& {Eisenstein}}{{Seo} \&
  {Eisenstein}}{2003}]{2003ApJ...598..720S}
{Seo} H.-J.,  {Eisenstein} D.~J.,  2003, \mn@doi [\apj] {10.1086/379122}, \href
  {https://ui.adsabs.harvard.edu/abs/2003ApJ...598..720S} {598, 720}

\bibitem[\protect\citeauthoryear{{Shandarin} \& {Zeldovich}}{{Shandarin} \&
  {Zeldovich}}{1989}]{1989RvMP...61..185S}
{Shandarin} S.~F.,  {Zeldovich} Y.~B.,  1989, \mn@doi [Reviews of Modern
  Physics] {10.1103/RevModPhys.61.185}, \href
  {https://ui.adsabs.harvard.edu/abs/1989RvMP...61..185S} {61, 185}

\bibitem[\protect\citeauthoryear{{Sheth} \& {Diaferio}}{{Sheth} \&
  {Diaferio}}{2001}]{2001MNRAS.322..901S}
{Sheth} R.~K.,  {Diaferio} A.,  2001, \mn@doi [\mnras]
  {10.1046/j.1365-8711.2001.04202.x}, \href
  {https://ui.adsabs.harvard.edu/abs/2001MNRAS.322..901S} {322, 901}

\bibitem[\protect\citeauthoryear{{Sheth} \& {Tormen}}{{Sheth} \&
  {Tormen}}{1999}]{1999MNRAS.308..119S}
{Sheth} R.~K.,  {Tormen} G.,  1999, \mn@doi [\mnras]
  {10.1046/j.1365-8711.1999.02692.x}, \href
  {https://ui.adsabs.harvard.edu/abs/1999MNRAS.308..119S} {308, 119}

\bibitem[\protect\citeauthoryear{{Snider}}{{Snider}}{1972}]{1972PhRvL..28..853S}
{Snider} J.~L.,  1972, \mn@doi [\prl] {10.1103/PhysRevLett.28.853}, \href
  {https://ui.adsabs.harvard.edu/abs/1972PhRvL..28..853S} {28, 853}

\bibitem[\protect\citeauthoryear{{Takada} et~al.,}{{Takada}
  et~al.}{2014}]{2014PASJ...66R...1T}
{Takada} M.,  et~al., 2014, \mn@doi [\pasj] {10.1093/pasj/pst019}, \href
  {https://ui.adsabs.harvard.edu/abs/2014PASJ...66R...1T} {66, R1}

\bibitem[\protect\citeauthoryear{{Tansella}, {Bonvin}, {Durrer}, {Ghosh}  \&
  {Sellentin}}{{Tansella} et~al.}{2018}]{2018JCAP...03..019T}
{Tansella} V.,  {Bonvin} C.,  {Durrer} R.,  {Ghosh} B.,   {Sellentin} E.,
  2018, \mn@doi [\jcap] {10.1088/1475-7516/2018/03/019}, \href
  {https://ui.adsabs.harvard.edu/abs/2018JCAP...03..019T} {2018, 019}

\bibitem[\protect\citeauthoryear{{Taruya} \& {Okumura}}{{Taruya} \&
  {Okumura}}{2020}]{2020ApJ...891L..42T}
{Taruya} A.,  {Okumura} T.,  2020, \mn@doi [\apjl] {10.3847/2041-8213/ab7934},
  \href {https://ui.adsabs.harvard.edu/abs/2020ApJ...891L..42T} {891, L42}

\bibitem[\protect\citeauthoryear{{Taruya}, {Saito}  \& {Nishimichi}}{{Taruya}
  et~al.}{2011}]{2011PhRvD..83j3527T}
{Taruya} A.,  {Saito} S.,   {Nishimichi} T.,  2011, \mn@doi [\prd]
  {10.1103/PhysRevD.83.103527}, \href
  {https://ui.adsabs.harvard.edu/abs/2011PhRvD..83j3527T} {83, 103527}

\bibitem[\protect\citeauthoryear{{Taruya}, {Saga}, {Breton}, {Rasera}  \&
  {Fujita}}{{Taruya} et~al.}{2020}]{2020MNRAS.491.4162T}
{Taruya} A.,  {Saga} S.,  {Breton} M.-A.,  {Rasera} Y.,   {Fujita} T.,  2020,
  \mn@doi [\mnras] {10.1093/mnras/stz3272}, \href
  {https://ui.adsabs.harvard.edu/abs/2020MNRAS.491.4162T} {491, 4162}

\bibitem[\protect\citeauthoryear{{Turneaure}, {Will}, {Farrell}, {Mattison}  \&
  {Vessot}}{{Turneaure} et~al.}{1983}]{1983PhRvD..27.1705T}
{Turneaure} J.~P.,  {Will} C.~M.,  {Farrell} B.~F.,  {Mattison} E.~M.,
  {Vessot} R.~F.~C.,  1983, \mn@doi [\prd] {10.1103/PhysRevD.27.1705}, \href
  {https://ui.adsabs.harvard.edu/abs/1983PhRvD..27.1705T} {27, 1705}

\bibitem[\protect\citeauthoryear{{Vessot} \& {Levine}}{{Vessot} \&
  {Levine}}{1979}]{1979GReGr..10..181V}
{Vessot} R.~F.~C.,  {Levine} M.~W.,  1979, \mn@doi [General Relativity and
  Gravitation] {10.1007/BF00759854}, \href
  {https://ui.adsabs.harvard.edu/abs/1979GReGr..10..181V} {10, 181}

\bibitem[\protect\citeauthoryear{{Vessot} et~al.,}{{Vessot}
  et~al.}{1980}]{1980PhRvL..45.2081V}
{Vessot} R.~F.~C.,  et~al., 1980, \mn@doi [\prl] {10.1103/PhysRevLett.45.2081},
  \href {https://ui.adsabs.harvard.edu/abs/1980PhRvL..45.2081V} {45, 2081}

\bibitem[\protect\citeauthoryear{{Werle} et~al.,}{{Werle}
  et~al.}{2022}]{2022ApJ...930...43W}
{Werle} A.,  et~al., 2022, \mn@doi [\apj] {10.3847/1538-4357/ac5f06}, \href
  {https://ui.adsabs.harvard.edu/abs/2022ApJ...930...43W} {930, 43}

\bibitem[\protect\citeauthoryear{Will}{Will}{2018}]{will_2018}
Will C.~M.,  2018, Theory and Experiment in Gravitational Physics, 2 edn.
Cambridge University Press, \mn@doi{10.1017/9781316338612}

\bibitem[\protect\citeauthoryear{{Wojtak}, {Hansen}  \& {Hjorth}}{{Wojtak}
  et~al.}{2011}]{2011Natur.477..567W}
{Wojtak} R.,  {Hansen} S.~H.,   {Hjorth} J.,  2011, \mn@doi [\nat]
  {10.1038/nature10445}, \href
  {https://ui.adsabs.harvard.edu/abs/2011Natur.477..567W} {477, 567}

\bibitem[\protect\citeauthoryear{{Yan}, {Raza}, {Van Waerbeke}, {Mead},
  {McCarthy}, {Tr{\"o}ster}  \& {Hinshaw}}{{Yan}
  et~al.}{2020}]{2020MNRAS.493.1120Y}
{Yan} Z.,  {Raza} N.,  {Van Waerbeke} L.,  {Mead} A.~J.,  {McCarthy} I.~G.,
  {Tr{\"o}ster} T.,   {Hinshaw} G.,  2020, \mn@doi [\mnras]
  {10.1093/mnras/staa295}, \href
  {https://ui.adsabs.harvard.edu/abs/2020MNRAS.493.1120Y} {493, 1120}

\bibitem[\protect\citeauthoryear{{Yoo}}{{Yoo}}{2010}]{2010PhRvD..82h3508Y}
{Yoo} J.,  2010, \mn@doi [\prd] {10.1103/PhysRevD.82.083508}, \href
  {https://ui.adsabs.harvard.edu/abs/2010PhRvD..82h3508Y} {82, 083508}

\bibitem[\protect\citeauthoryear{{Yoo}}{{Yoo}}{2014}]{2014CQGra..31w4001Y}
{Yoo} J.,  2014, \mn@doi [Classical and Quantum Gravity]
  {10.1088/0264-9381/31/23/234001}, \href
  {https://ui.adsabs.harvard.edu/abs/2014CQGra..31w4001Y} {31, 234001}

\bibitem[\protect\citeauthoryear{{Yoo} \& {Zaldarriaga}}{{Yoo} \&
  {Zaldarriaga}}{2014}]{2014PhRvD..90b3513Y}
{Yoo} J.,  {Zaldarriaga} M.,  2014, \mn@doi [\prd]
  {10.1103/PhysRevD.90.023513}, \href
  {https://ui.adsabs.harvard.edu/abs/2014PhRvD..90b3513Y} {90, 023513}

\bibitem[\protect\citeauthoryear{{Zel'dovich}}{{Zel'dovich}}{1970}]{1970A&A.....5...84Z}
{Zel'dovich} Y.~B.,  1970, \aap, \href
  {https://ui.adsabs.harvard.edu/abs/1970A%26A.....5...84Z} {5, 84}

\bibitem[\protect\citeauthoryear{{Zhao}, {Peacock}  \& {Li}}{{Zhao}
  et~al.}{2013}]{2013PhRvD..88d3013Z}
{Zhao} H.,  {Peacock} J.~A.,   {Li} B.,  2013, \mn@doi [\prd]
  {10.1103/PhysRevD.88.043013}, \href
  {https://ui.adsabs.harvard.edu/abs/2013PhRvD..88d3013Z} {88, 043013}

\bibitem[\protect\citeauthoryear{{Zhu}, {Alam}, {Croft}, {Ho}  \&
  {Giusarma}}{{Zhu} et~al.}{2017}]{2017MNRAS.471.2345Z}
{Zhu} H.,  {Alam} S.,  {Croft} R. A.~C.,  {Ho} S.,   {Giusarma} E.,  2017,
  \mn@doi [\mnras] {10.1093/mnras/stx1644}, \href
  {https://ui.adsabs.harvard.edu/abs/2017MNRAS.471.2345Z} {471, 2345}

\makeatother
\end{thebibliography}

\appendix

\section{Derivation of Equation (3)}
\label{sec: app Eq3}

We present the derivation of the analytical model for the dipole cross-correlation function given at Eq.~(\ref{eq: xi1}). This derivation builds on our previous works~\citep{2020MNRAS.498..981S,2022MNRAS.511.2732S}, but we here succinctly summarize our treatment and the approximation, which enable us to compute the dipole cross-correlation with one-dimensional numerical integration.
After presenting a rigorous expression for the observed density field based on the Zel'dovich approximation in Appendix~\ref{sec: app A1}, a simplified expression is derived in Appendix~\ref{sec: app A2}, leading to an analytical model of the dipole cross-correlation involving only the one-dimensional integrals.

\subsection{Preliminary}
\label{sec: app A1}

Our starting point is a mapping relation between the real and redshift space given at Eq.~(\ref{eq: s to x}) in the main text, which includes the standard Doppler effect, the gravitational redshift effect due to the non-linear halo potential, and contributions arising from the velocity dispersions. In what follows,  we denote the relativistic effects, except the standard Doppler term, by $\epsilon_{\rm NL}$. Eq.~(\ref{eq: s to x}) is then rewritten with
\begin{align}
\bm{s} &= \bm{x} + \frac{1}{aH}\left( \bm{v}\cdot \hat{\bm{x}}\right)\hat{\bm{x}} + \epsilon_{\rm NL}\hat{\bm{x}} , \label{eq: app s to x}
\\
\epsilon_{\rm NL} &= 
- \frac{1}{aH}\left( \phi_{\rm halo} - \gamma v^{2}_{\rm g} \right)
, \label{eq: epsilon NL}
\end{align}
where $\phi_{\rm halo}$ and $v^{2}_{\rm g}$ stand for, respectively, the gravitational potential of haloes and velocity dispersions, whose explicit modelling is presented in Appendix~\ref{sec: app model NL}.
In Eq.~(\ref{eq: epsilon NL}), the first and second terms in the right-hand side express the gravitational redshift effect arising from the deep halo potential and the sum of the transverse Doppler, light-cone, and surface brightness modulation effects mainly due to the velocity dispersions of galaxies, which are the first and second major contributions to the small-scale dipole signal, respectively~\citep{2019MNRAS.483.2671B}.
Note that we set $\gamma=-5/2$ as a fiducial value, taken from \citet{2013MNRAS.435.1278K}. Here, the quantity $\epsilon_{\rm NL}$ is described by the non-perturbative contribution due to the nonlinearies of the halo/galaxy formation and evolution.

In measuring the dipole cross-correlation between different types of galaxies, we are particularly interested in a pair of galaxy samples, each of which resides at similar halos having mostly the same value of $\epsilon_{\rm NL}$. This correlation can be regarded as the conditional average over the halo density field for a fixed $\epsilon_{\rm NL}$.
Hence, we treat the quantity $\epsilon_{\rm NL}$ not as a random variable but as a constant value which depends on the halo mass and redshift.
Although a more general expression of the mapping relation discussed in the literature involves contributions from the integrated Sachs-Wolfe and Shapiro time-delay effects \citep[e.g.,][]{2010PhRvD..82h3508Y,2011PhRvD..84d3516C,2011PhRvD..84f3505B}, we retain relevant contributions in $\epsilon_{\rm NL}$ to the scales of our interest, $\lesssim30\,{\rm Mpc}/h$~\citep{2019MNRAS.483.2671B,2020MNRAS.498..981S}.

Following our previous works~\citep{2020MNRAS.491.4162T,2020MNRAS.498..981S}, we adopt the Zel'dovich approximation, known as the first-order Lagrangian perturbation theory~\citep{1969JETP...30..512N,1970A&A.....5...84Z,1989RvMP...61..185S}, in order, which provides a simple way to deal with redshift-space distortions involving the wide-angle effect even beyond linear regime.
The building block in Lagrangian perturbation theory is a displacement field, which relates the Eulerian position, $\bm{x}$, to the Lagrangian position (initial position), $\bm{q}$, at the time of interest.
Denoting the displacement field in the Zel'dovich approximation by $\bm{\Psi}(\bm{q},t)$, which is related to the Lagrangian linear density field $\delta_{\rm L}(\bm{q},t)$ through $\bm{\nabla}\cdot\bm{\Psi}(\bm{q},t)=-\delta_{\rm L}(\bm{q},t)$, and assuming that the objects of our interest follow the velocity flow of mass distributions (no velocity bias), we express the Eulerian position $\bm{x}$ and velocity field $\bm{v}$ as
\begin{align}
\bm{x}(\bm{q},t) &= \bm{q} + \bm{\Psi}(\bm{q},t) , \label{eq: x to q}\\
\bm{v}(\bm{q},t) &= a \frac{{\rm d}\bm{\Psi}(\bm{q},t)}{{\rm d}t} = aHf\bm{\Psi}(\bm{q},t) , \label{eq: Zel velocity}
\end{align}
where we define the linear growth rate by $f \equiv {\rm d}\ln{D_{+}(a)}/{\rm d}\ln{a}$ with $D_{+}(t)$ being the linear growth factor.
The second equality is valid in the Zel'dovich approximation, where the time dependence of the displacement field is solely encapsulated in the factor $D_+(t)$, i.e., $\bm{\Psi}\propto D_+(t)$.

Substituting Eqs.(\ref{eq: x to q}) and (\ref{eq: Zel velocity}) into Eq.~(\ref{eq: app s to x}), the relation between the redshift-space position, $\bm{s}$, and the Lagrangian position, $\bm{q}$, becomes
\begin{align}
s_{i} &= q_i + \left( \delta_{ij}+f\,\hat{x}_i\hat{x}_j \right) \Psi_i(\bm{q})+\epsilon(\bm{x})\hat{x}_i
\nonumber
\\
&\simeq q_{i} + R_{ij}(\hat{\bm{q}})\Psi_{j}(\bm{q}) + \epsilon(\bm{q}) \hat{q}_{i},
\label{eq: mapping_s_q-space epsilon}
\end{align}
where we define $R_{ij}(\hat{\bm{q}}) \equiv \delta_{ij} + f\hat{q}_{i}\hat{q}_{j}$. Here and hereafter, we use the Einstein summation convention. Note that the second line is valid in the Zel'dovich approximation.

Using the number conservation between Lagrangian space and redshift space, we express the number density of the biased objects X in redshift space, $n^{\rm(S)}_{\rm X}(\bm{s})$, in terms of the Lagrangian space quantities through Eq.~(\ref{eq: mapping_s_q-space epsilon}):
\begin{align}
n^{({\rm S})}_{\rm X}(\bm{s})
&=
\overline{n}_{\rm X} \left( 1 + b^{\rm L}_{\rm X} \delta_{\rm L}(\bm{q})\right) \left| \frac{\partial s_{i}}{\partial q_{j}}\right|^{-1}
\notag \\
&= 
\overline{n}_{\rm X} \int{\rm d}^{3}\bm{q}\, \left( 1 + b^{\rm L}_{\rm X} \delta_{\rm L}(\bm{q})\right) \delta_{\rm D}(s_{i} - q_{i} -R_{ij}\Psi_{j} + \epsilon \hat{q}_{i})
\notag \\
&= \overline{n}_{\rm X}\int{\rm d}^{3}\bm{q}\; \int\frac{{\rm d}^{3}\bm{k}}{(2\pi)^{3}}\;
{\rm e}^{{\rm i} k_{i}\left( s_{i}-q_{i}-R_{ij}\Psi_{j} - \epsilon\hat{q}_{i}\right)} 
\left( 1 + b^{\rm L}_{\rm X} \delta_{\rm L}(\bm{q})\right) . \label{eq: def n^{S}}
\end{align}
Here, we assume the linear galaxy bias, and the quantity $b^{\rm L}_{\rm X}$ is the Lagrangian linear bias parameter of the biased objects X, whichis related to the Eulerian bias $b_{\rm X}$ through $b_{\rm X} = 1 + b^{\rm L}_{\rm X}$. The quantity  $\overline{n}_{\rm X}$ is the mean number density of the biased objects X at a given redshift.

Provided the number density, the density fluctuation $\delta^{\rm (S)}_{\rm X}$ is defined as follows
\begin{align}
\delta^{\rm (S)}_{\rm X}(\bm{s}) \equiv \frac{n^{({\rm S})}_{\rm X}(\bm{s})}{\Braket{n^{({\rm S})}_{\rm X}(\bm{s})}} -1 ,
\label{eq: def delta}
\end{align}
where the bracket $\Braket{\cdots}$ stands for the ensemble average.
We note that the quantity $\langle n^{\rm(S)}_{\rm X}\rangle$ differs from $\overline{n}_{\rm X}$, due to the directional-dependence $\hat{\bm{q}}$ in $R_{ij}(\hat{\bm{q}})$ and $\epsilon(\bm{q}) \hat{\bm{q}}$~\citep[see][]{2020MNRAS.491.4162T,2020MNRAS.498..981S}.

With the definition given by Eq.~(\ref{eq: def delta}), an analytical expression of the correlation function is regorously derived in \citet{2020MNRAS.498..981S}, without invoking other approximations except the Zel'dovich approximation under the Gaussianity of the linear density field $\delta_{\rm L}$ and linear galaxy bias (see Eqs.~(3.14)--(3.18) in their paper).

\subsection{Analytical model of dipole cross-correlation}
\label{sec: app A2}

Here, we derive a simplied analytical expression of dipole cross correlation function, following \citet{2022MNRAS.511.2732S}. First, we linearise the expression at Eq.~(\ref{eq: def delta}) with respect to the displacement field. \citet{2020MNRAS.491.4162T} showed that this treatment still gives an accurate description of the dipole cross-correlation function at the scales of our interest. On top of this, we also expand the term proportional to $\epsilon_{\rm NL}$ from the exponent, which are supposed to be small though the term involves non-perturbative contributions.
Then, we obtain
\begin{align}
\delta^{({\rm S})}(\bm{s}) &= \int{\rm d}^{3}\bm{q}\; \int\frac{{\rm d}^{3}\bm{k}}{(2\pi)^{3}}\;
{\rm e}^{{\rm i}\bm{k}\cdot\left( \bm{s}-\bm{q}\right)}
\Biggl[
b^{\rm L}\delta_{\rm L} - {\rm i}k_{i} R_{ij}\Psi_{j}
\notag \\
& \qquad 
+ \epsilon_{\rm NL} \left( - ({\rm i}\bm{k}\cdot\hat{\bm{q}}) + \frac{2}{s} \right)
\left( b^{\rm L}\delta_{\rm L} - {\rm i}k_{i} R_{ij}\Psi_{j} \right)
\Biggr] ,
\label{eq: delta s pre int 2}
\end{align}
where the first line consists of the the real-space contribution and standard Doppler term, while the second line stands for the leading-order contribtions of the non-perturbative relativistic correction.

Now we define the correlation function between the populations X at $\bm{s}_{1}$ and Y at $\bm{s}_{2}$:
\begin{align}
\xi(\bm{s}_1,\bm{s}_2) \equiv \Braket{\delta^{(\rm S)}_{\rm X}(\bm{s}_{1}) \delta^{(\rm S)}_{\rm Y}(\bm{s}_{2})} .
\label{eq: xi def}
\end{align}
Substituting Eq.~(\ref{eq: delta s pre int 2}) into Eq.~(\ref{eq: xi def}), and performing the $\bm{q}$-integrals, we obtain the expression for the cross-correlation function~\citep{2022MNRAS.511.2732S}:
\begin{align}%
\xi(\bm{s}_{1},\bm{s}_{2}) & =
\int\frac{{\rm d}^{3}k}{(2\pi)^{3}}{\rm e}^{{\rm i}\bm{k}\cdot\bm{s}}
P_{\rm L}(k)
\\
& \times
\Biggl[
\left( b_{\rm X} + f\mu^{2}_{k1} + {\rm i} f \frac{2}{ks_{1}}\mu_{k1}\right)
\left( b_{\rm Y} + f\mu^{2}_{k2} - {\rm i} f \frac{2}{ks_{2}}\mu_{k2} \right)
\notag \\
& +
\frac{\epsilon_{\rm NL, X}}{s_{1}}
\Biggl(
-1 + \mu^{2}_{k1} + {\rm i}f\frac{2}{ks_{1}}\mu_{k1}
+ {\rm i} b_{\rm X} ks_{1} \mu_{k1}
\notag \\
& 
-2f\mu^{2}_{k1} + {\rm i} \frac{2}{ks_{1}}\mu_{k1} + {\rm i} f ks_{1} \mu^{3}_{k1}
\Biggr)
\Biggl(
b_{\rm Y} + f\mu^{2}_{k2} - {\rm i} f \frac{2}{ks_{2}}\mu_{k2}
\Biggr)
\notag \\
&+
\frac{\epsilon_{\rm NL, Y}}{s_{2}}
\Biggl(
-1 + \mu^{2}_{k2} - {\rm i} f\frac{2}{ks_{2}}\mu_{k2}
- {\rm i} b_{\rm Y} ks_{2} \mu_{k2}-2f\mu^{2}_{k2}
\notag \\
&
- {\rm i}\frac{2}{ks_{2}}\mu_{k2} - {\rm i} f ks_{2} \mu^{3}_{k2}
\Biggr)
\Biggl(
b_{\rm X} + f\mu^{2}_{k1} + {\rm i} f \frac{2}{ks_{1}}\mu_{k1}
\Biggr)
\Biggr]
, \label{eq: delta s pre int 3}
\end{align}
where we define $\mu_{k1} = \hat{\bm{s}}_{1}\cdot\hat{\bm{k}}$ and $\mu_{k2} = \hat{\bm{s}}_{2}\cdot\hat{\bm{k}}$. The function $P_{\rm L}(k)$ stands for the linear power spectrum of the density field $\delta_{\rm L}$ given by
\begin{align}
\Braket{\delta_{\rm L}(\bm{k})\delta_{\rm L}(\bm{k}')} = (2\pi)^{3}\delta_{\rm D}(\bm{k}+\bm{k}')P_{\rm L}(k) . \label{eq: def PL(k)}
\end{align}

In Eq.~(\ref{eq: delta s pre int 3}), introducing a polar coordinate system in $\bm{k}$, the integral over the azmuthal angle performed analytically using the formulae given in Appendix A in \citet{2022MNRAS.511.2732S}.
Then, the dependence of the correlation function on the vectors $\bm{s}_{1}$ and $\bm{s}_{2}$ in Eq.~(\ref{eq: delta s pre int 3}) is described by only the following quantities: $(\hat{\bm{s}}\cdot\hat{\bm{s}}_{1})$, $(\hat{\bm{s}}\cdot\hat{\bm{s}}_{2})$, $(\hat{\bm{s}}_{1}\cdot\hat{\bm{s}}_{2})$, $s_{1}$, and $s_{2}$. These quantities can be rewritten in terms of the following three variables, i.e., separation $s = |\bm{s}_{2} - \bm{s}_{1}|$, the line-of-sight distance $d = |\bm{s}_{1} + \bm{s}_{2}|/2$, and directional cosine $\mu = \hat{\bm{s}} \cdot \hat{\bm{d}}$. Since we are interested in the cases with $s\ll d$, we can expand the quantities as   
\begin{equation}
s_{1,2} \simeq d\left( 1 \mp \frac{1}{2}\frac{s}{d} \mu\right) 
,~~~
\hat{\bm{s}}\cdot\hat{\bm{s}}_{1,2}  \simeq \mu \mp \frac{1}{2}(1-\mu^{2})\frac{s}{d}
,~~~
\hat{\bm{s}}_{1}\cdot\hat{\bm{s}}_{2}  \simeq 1
,
\end{equation}
with $-$ for $\bm{s}_{1}$ and $+$ for $\bm{s}_{2}$, and these relations are valid at $\mathcal{O}(s/d)$.
Then, the resultant expression is given as a polynomial form of $\mu$.
After performing the multipole expansion of $\xi$ in terms of $\mu$:
\begin{align}
\xi_{1}(s,d)
& = \frac{3}{2}\int^{1}_{-1}{\rm d}\mu\; \xi_{{\rm XY}}(s,d,\mu) \, \mu ,
\label{eq: multipole expansion}
\end{align}
we arrive straightforwardly at Eq.~(\ref{eq: xi1}) in the main text:
\begin{align}
\xi_{1}(s,d) &=
2f\Delta b\frac{s}{d}\left( \Xi^{(1)}_{1}(s) - \frac{\Xi^{(0)}_{2}(s)}{5} \right)
\notag \\
& -
\frac{\Delta \epsilon_{\rm NL}}{saH} \left( b_{\rm X}b_{\rm Y} + \frac{3}{5}(b_{\rm X}+b_{\rm Y})f + \frac{3}{7}f^{2}\right) \Xi^{(-1)}_{1}(s)
, \label{eq: app xi1}
\end{align}
where $\Delta b  = b_{\rm X} - b_{\rm Y}$ and $\Delta \epsilon_{\rm NL} = \epsilon_{\rm NL, X} - \epsilon_{\rm NL, Y}$, and we define 
\begin{align}
\Xi^{(n)}_{\ell}(s) = \int\, \frac{k^{2} {\rm d}k}{2\pi^{2}} \frac{j_{\ell}(ks)}{(ks)^{n}} P_{\rm L}(k) .
\end{align}
In Eq.~(\ref{eq: app xi1}), the first and second lines are, respectively, proportional to $\delta^{2}_{\rm L}$ and $\phi_{\rm halo,X/Y} \delta^{2}_{\rm L}$ or $v^{2}_{\rm g, X/Y} \delta^{2}_{\rm L}$. This dependence comes from the fact the we retain the most dominant leading-order contribution to reproduce the prediction by the exact expression. Hence, Eq.~(\ref{eq: app xi1}) is still relevant to describe major relativistic effects, similar to those involving a more intricate second-order expressions in perturbation theory ~\citep[e.g.,][]{2014JCAP...11..013B,2014JCAP...12..017D,2014PhRvD..90b3513Y}.

It has been shown in \citet{2022MNRAS.511.2732S} that the dipole cross-correlation based on a rigorous calculation of Eqs.~(\ref{eq: def n^{S}}) and (\ref{eq: def delta}) (Eqs.~(3.14)--(3.18) of their paper) agrees very well with the prediction by Eq.~(\ref{eq: app xi1}), even down to  $s\approx 5\, {\rm Mpc}/h$. This ensures that an accurate calculation of Fisher matrix is possible with the model given by Eq.~(\ref{eq: app xi1}), allowing systematic investigations with less computationally cost.

\section{Model of the non-perturbative terms}
\label{sec: app model NL}

Here, we present the model of the non-perturbative term in Eq.~(\ref{eq: epsilon NL}). The non-perturbative term contains two contributions. One is the gravitational redshift effect due to the gravitational potential of haloes and another is the contribution from the velocity dispersion of galaxies, which are presented in Appendices~\ref{sec: app B1} and \ref{sec: app B2}, respectively.

\subsection{Halo gravitational potential}
\label{sec: app B1}

Based on the the universal halo density profile called Navarro-Frenk-White (NFW) profile~\citep{1996ApJ...462..563N} and its gravitational potential, we present an analytical model for the non-perturbative halo potential, $\phi_{\rm halo}$, given in Eq.~(\ref{eq: epsilon NL}).
The NFW profile quantitatively describes the halo density profiles in cosmological $N$-body simulations, given by
\begin{equation}
\rho_{\rm NFW}(r, z, M) =
\frac{\rho_{\rm s}(z, M)}
{\left( r/r_{\rm s}(z, M) \right)
\left\{ 1+\left( r/r_{\rm s}(z, M) \right) \right\}^{2}} ,
\label{eq: NFW profile}
\end{equation}
with $r$, $z$, and $M$ being radius from the halo centre, redshift, and halo mass, respectively. The overdensity, $\rho_{\rm s}(z,M)$, and the scale radius, $r_{\rm s}(z,M)$, are related to the concentration parameter $c_{\rm vir}$ through
\begin{align}
\rho_{\rm s}(z, M) &= \frac{\Delta_{\rm vir}(z)\rho_{\rm m0}}{3}c^{3}_{\rm vir}(M, z) 
\notag \\
&\times 
\left[ \ln{\left( 1 +c_{\rm vir}(z, M) \right)} - \frac{c_{\rm vir}(z, M)}{1+c_{\rm vir}(z, M)} \right]^{-1}
, \\
r_{\rm s}(z, M) &= \frac{r_{\rm vir}(z, M)}{c_{\rm vir}(z, M)} , 
\end{align}
where we define the virial radius $r_{\rm vir}$ and virial overdensity $\Delta_{\rm vir}$ by~\citep{1998ApJ...495...80B,2001MNRAS.321..559B}
\begin{align}
r_{\rm vir}(z, M) &= \left( \frac{3M}{4\pi\Delta_{\rm vir}(z)\rho_{\rm m0}} \right)^{\frac{1}{3}} , \\
\Delta_{\rm vir}(z) &= \frac{18\pi^{2} + 82\left( \Omega_{\rm m}(z) -1\right) - 39\left( \Omega_{\rm m}(z) -1\right)^{2}}{\Omega_{\rm m}(z)},\\
\Omega_{\rm m}(z) &= \frac{(1+z)^{3}\Omega_{\rm m0}}{(1+z)^{3}\Omega_{\rm m0} + \Omega_{\Lambda 0}} .
\end{align}
We use the following fitting form for the concentration parameter~\citep{2001MNRAS.321..559B,2002PhR...372....1C}: 
\begin{equation}
c_{\rm vir}(z, M) = \frac{9}{1+z}\left( \frac{M}{M_{*}(z)}\right)^{-0.13} ,
\end{equation}
where $M_{*}(z)$ stands for the characteristic mass scale defined by $\sigma_{M}(M_{*})D_{+}(z) = \delta_{\rm crit}$. The quantity $\delta_{\rm crit}$ is the critical over-density of the spherical collapse model, and $\sigma_{M}$ is the root-mean square amplitude of the matter density fluctuations smoothed with top-hat filter of the radius $R = \left( 3M/(4\pi \overline{\rho})\right)^{1/3}$ with $\overline{\rho}$ being the mean mass density.

Solving the Poisson equation, we obtain the gravitational potential of the NFW profile at Eq.~(\ref{eq: NFW profile}) under the boundary condition $\phi_{\rm NFW}\to0$ at $r\to\infty$:
\begin{align}
\phi_{\rm NFW}(r, z, M) &= -4\pi G (1+z) \rho_{\rm s}(z, M) r^{2}_{\rm s}(z, M)
\notag \\
& \times
\left(\frac{r}{r_{\rm s}(z, M)}\right)^{-1}\ln{\left( 1 + \frac{r}{r_{\rm s}(z, M)}\right)} . \label{eq: phi NFW(r)}
\end{align}

To incorporate the impact from the off-centered galaxy position into the potential estimate, we introduce the probability distribution function of the galaxy position inside each halo, $p_{\rm off}$, normalized as follows~\citep{2013MNRAS.435.2345H}: 
\begin{align}
\int^{r_{\rm vir}}_{0}\, 4\pi r^{2}p_{\rm off}(r; R_{\rm off})\, {\rm d}r = 1 .
\label{eq: norm poff}
\end{align}
The probability distribution function is assumed to be the Gaussian distribution, i.e., $p_{\rm off}(r; R_{\rm off}) \propto \exp{\left( -(r/R_{\rm off})^{2}/2\right)}$ with $R_{\rm off}$ being the offset parameter. Using the distribution function $p_{\rm off}$, we estimate the halo potential at the off-centered galaxy position by
\begin{align}
\overline{\phi}_{\rm NFW}(z,M,R_{\rm off}) &= \int^{r_{\rm vir}}_{0}\, 4\pi r^{2}\phi_{\rm NFW}(r,z,M)p_{\rm off}(r;R_{\rm off})\, {\rm d}r , \label{eq: average OC}
\end{align}

Finally, the averaged NFW potential is adopted as the model $\phi_{\rm halo}$ in Eq.~(\ref{eq: epsilon NL}):
\begin{align}
\phi_{\rm halo} = \overline{\phi}_{\rm NFW}(z,M,R_{\rm off}) .
\end{align}
We note that the off-centering parameter $R_{\rm off}$ can be estimated by using even multipole moments independently of the dipole moment~\citep{2013MNRAS.435.2345H}.

\subsection{Velocity dispersions}
\label{sec: app B2}

Next, we consider the velocity dispersion contribution, $v^{2}_{\rm g}$, in Eq.~(\ref{eq: epsilon NL}), which is expressed as a sum of the two contributions~\citep[e.g.,][]{2001MNRAS.322..901S}:
\begin{align}
v^{2}(r,z,M) = v^{2}_{\rm vir}(r,z,M) + v^{2}_{\rm halo}(z,M)
. \label{eq: v2 sum}
\end{align}
Here, the first and second terms at the right-hand side are originated respectively from the virial motion within a halo and the large-scale coherent motion of the host haloes. The second term, which is a subdominant contribution, is non-vanishing even if the galaxies reside at the centre of the haloes.

To compute the velocity dispersion of the virial motion, $v^{2}_{\rm vir}$, we adopt the analytical formula for the velocity dispersion of the NFW density profile~\citep[see Eq.~(14) of ][]{2001MNRAS.321..155L}:
\begin{align}
v^{2}_{\rm vir}(r,z,M) = \alpha(r,z,M)\frac{GM}{r_{\rm vir}}
, \label{eq: v2 vir}
\end{align}
with the function $\alpha(r,z,M)$ given by
\begin{align}
\alpha(r,z,M) &= \frac{3}{2}c_{\rm vir}^{2} g(c_{\rm vir})x(1+c_{\rm vir}x)^{2}\Biggl[
6{\rm Li}_{2}(-c_{\rm vir}x) + \pi^{2} -\ln{(c_{\rm vir}x)}
\notag \\
& \quad 
 - \frac{1}{c_{\rm vir}x}
- \frac{1}{(1+c_{\rm vir}x)^{2}} - \frac{6}{1+c_{\rm vir}x} +3\ln^{2}(1+c_{\rm vir}x)
\notag \\
& \quad 
+ \ln{(1+c_{\rm vir}x)}\left( 1 + \frac{1}{(c_{\rm vir}x)^{2}} - \frac{4}{c_{\rm vir}x} - \frac{2}{1+ c_{\rm vir}x}\right)
\Biggr]
, \label{eq: alpha}
\end{align}
where the quantity $x$ and function ${\rm Li}_{2}(x)$ respectively stand for the radius normalized by the virial radius, $x \equiv r/r_{\rm vir}$, and the dilogarithm. The function $g(c_{\rm vir})$ is defined as $g(c_{\rm vir}) \equiv \left[ \ln(1+c_{\rm vir})-c_{\rm vir}/(1+c_{\rm vir})\right]^{-1}$.

We estimate the velocity dispersion due to the large-scale coherent motion, $v^{2}_{\rm halo}$, by using the peak theory prediction based on the linear Gaussian density fields~\citep[][]{1986ApJ...304...15B,2001MNRAS.322..901S}:
\begin{align}
v^{2}_{\rm halo}(z,M) = (aHfD_{+})^{2}\sigma^{2}_{-1}(M)\left( 1-\frac{\sigma^{4}_{0}(M)}{\sigma^{2}_{1}(M)\sigma^{2}_{-1}(M)} \right)
, \label{eq: v2 halo}
\end{align}
where we define the function $\sigma_{n}$ by
\begin{align}
\sigma^{2}_{n}(M) = \int\frac{k^{2}{\rm d}k}{2\pi^{2}}\, k^{2n}P_{\rm L}(k)W^{2}(kR) .
\end{align}
Here, the function $W(x) = 3j_{1}(x)/x$ is the Fourier transform of the real space top-hat window function, and the radius $R$ is related to the halo mass $M$ through $M = 4\pi \bar{\rho} R^{3}/3$ with $\bar{\rho}$ being the background matter density.

We take the off-centering effect into account the velocity dispersion Eq.~(\ref{eq: v2 sum}) by the same averaging procedure in Eq.~(\ref{eq: average OC}):
\begin{align}
\overline{v^{2}}(z,M,R_{\rm off}) &= \int^{r_{\rm vir}}_{0}\, 4\pi r^{2}v^{2}(r,z,M)p_{\rm off}(r;R_{\rm off})\, {\rm d}r , \label{eq: average TD}
\end{align}
Finally, we adopt the velocity dispersion obtained in this way as the model $v^{2}_{g}$ in Eq.~(\ref{eq: epsilon NL}):
\begin{align}
v^{2}_{\rm g} = \overline{v^{2}}(z,M,R_{\rm off}) .
\end{align}

\section{Fisher analysis for uncertainty in the bias parameter}
\label{sec: app Fisher}

In the Fisher analysis for deriving the constraint on $\alpha$, the uncertainty in halo masses $\sigma_{M}$, inferred from the uncertainty in the bias parameters $\sigma_{b}$, is introduced as a Gaussian prior. 
We here describe the Fisher analysis for deriving the uncertainty in the bias parameter $\sigma_{b}$ in detail.

We consider the observed anisotropies arising only from the standard Doppler effect, which plays a dominant role when constraining the bias parameter $b$ as well as the linear growth rate $f$ and relevant cosmological parameters. This is particularly the case if we use the even multipole moments of the clustering anisotropies at large scales~\citep[e.g.,][]{2014PhRvD..89h3535B,2019MNRAS.483.2671B}, and hence we safely ignore the other relativitsic effects. Taking the plane parallel limit, we have the linear model of the galaxy auto power spectrum in redshift space~\citep{1987MNRAS.227....1K,1992ApJ...385L...5H}:
\begin{align}
P_{\rm gg}(k, \mu_{k}) = \left( b + f \mu_{k}\right)^{2} P_{\rm L}(k) , \label{eq: Kaiser}
\end{align}
where we define $\mu_{k} = \hat{\bm{k}}\cdot\hat{\bm{z}}$ with the constant line of sight vector $\hat{\bm{z}}$.
On top of the clustering anisotropy induced by the Doppler effect given in Eq.~(\ref{eq: Kaiser}), the observed power spectrum in comoving space exhibits additional anisotropies induced by the Alcock-Paczynski effect~\citep{1979Natur.281..358A}, which is modelled by replacing the projected wavenumbers perpendicular and parallel to the line-of-sight direction, $k_{\perp}$ and $k_{\parallel}$ with $\left( d_{\rm A}/d_{\rm A,fid}\right)k_{\perp}$ and $\left( H/H_{\rm fid}\right)^{-1} k_{\parallel}$, respectively, and further by multiplying the factor $\left( H/H_{\rm fid}\right)\left( d_{\rm A}/d_{\rm A,fid}\right)^{-2}$ with the power spectrum~\citep[see e.g.,][]{2003ApJ...598..720S,2011PhRvD..83j3527T}. Here, we define the angular diameter distance $d_{\rm A}$, and the parameters $H_{\rm fid}$ and $d_{\rm A,fid}$ stand for the quantities calculated in the fiducial cosmological model.

As a result, the power spectrum in the model is characterized by four parameters $\bm{\theta} = (b, f, H/H_{\rm fid}, d_{\rm A}/d_{\rm A,fid})$.
Assuming a flat $\Lambda$CDM model determined by the seven-year WMAP results~\citep{2011ApJS..192...18K}, we perform the Fishser analysis to derive the parameter constraints given the survey parameters, i.e., the mean redshift slice $z$, survey volume $V$ and mean number density of galaxies $n_{\rm g}$. The Fisher matrix is given by~\citep[e.g.,][]{2020ApJ...891L..42T}:
\begin{align}
F_{ij} &= \frac{V}{8\pi^{2}}\int^{k_{\rm max}}_{k_{\rm min}}k^{2}\,{\rm d}k\, \int^{1}_{-1}{\rm d}\mu_{k}\,
\notag \\
& \quad \times
\frac{\partial P_{\rm gg}(k,\mu_{k})}{\partial \theta_{i}}
\frac{\partial P_{\rm gg}(k,\mu_{k})}{\partial \theta_{j}}
\left( P_{\rm gg}(k,\mu_{k}) + n^{-1}_{\rm g}\right)^{-2} ,
\label{eq: fisher b}
\end{align}
where the minimum and maximum wavenumbers used in the analysis are set to $k_{\rm min}=2\pi\,V^{-1/3}$ and $k_{\rm max}=0.1\,{\rm Mpc}/h$, respectively. The survey volume and mean number density of galaxies for each survey can be found in Appendix~E in \citet{2022MNRAS.511.2732S}.

In this way, we obtain the marginalised constraint on the bias parameters, $\sigma_{b} = \sqrt{\left( F^{-1} \right)_{11}}$, which is used for a  Gaussian prior to further  constrain the LPI violation parameter $\alpha$ in Eq.~(\ref{eq: fisher alpha}).

\bsp	
\label{lastpage}
\end{document}